\documentclass[twocolumn]{aastex63}
\usepackage{amsmath}

\usepackage{ulem}

\received{January 14, 2020}
\revised{May 17, 2020}
\accepted{May 18, 2020}
\shorttitle{\textit{NuSTAR} observations of two non-merging LIRGs}
\shortauthors{Yamada et al.}
\graphicspath{{./}{figures/}}

\begin{document}

\title{Nature of Compton-thick Active Galactic Nuclei in ``Non-merging'' 
Luminous Infrared Galaxies UGC~2608 and NGC~5135 Revealed 
with Broadband X-ray Spectroscopy}

\correspondingauthor{Satoshi Yamada}
\email{styamada@kusastro.kyoto-u.ac.jp}

\author[0000-0002-9754-3081]{Satoshi Yamada}
\affiliation{Department of Astronomy, Kyoto University, Kitashirakawa-Oiwake-cho, Sakyo-ku, Kyoto 606-8502, Japan}

\author[0000-0001-7821-6715]{Yoshihiro Ueda}
\affiliation{Department of Astronomy, Kyoto University, Kitashirakawa-Oiwake-cho, Sakyo-ku, Kyoto 606-8502, Japan}

\author[0000-0002-0114-5581]{Atsushi Tanimoto}
\affiliation{Department of Astronomy, Kyoto University, Kitashirakawa-Oiwake-cho, Sakyo-ku, Kyoto 606-8502, Japan}

\author{Saeko Oda}
\affiliation{Department of Astronomy, Kyoto University, Kitashirakawa-Oiwake-cho, Sakyo-ku, Kyoto 606-8502, Japan}

\author[0000-0001-6186-8792]{Masatoshi Imanishi}
\affiliation{National Astronomical Observatory of Japan, Osawa, Mitaka, Tokyo 181-8588, Japan}
\affiliation{Department of Astronomical Science, Graduate University for Advanced Studies (SOKENDAI), 2-21-1 Osawa, Mitaka, Tokyo 181-8588, Japan}

\author[0000-0002-3531-7863]{Yoshiki Toba}
\affiliation{Department of Astronomy, Kyoto University, Kitashirakawa-Oiwake-cho, Sakyo-ku, Kyoto 606-8502, Japan}
\affiliation{Academia Sinica Institute of Astronomy and Astrophysics, 11F of Astronomy-Mathematics Building, AS/NTU, No.1, Section 4, Roosevelt Road, Taipei 10617, Taiwan}
\affiliation{Research Center for Space and Cosmic Evolution, Ehime University, 2-5 Bunkyo-cho, Matsuyama, Ehime 790-8577, Japan}

\author[0000-0001-5231-2645]{Claudio Ricci}
\affiliation{N\'ucleo de Astronom\'{\i}a de la Facultad de Ingenier\'{\i}a, Universidad Diego Portales, Av. Ej\'ercito Libertador 441, Santiago, Chile}
\affiliation{Kavli Institute for Astronomy and Astrophysics, Peking University, Beijing 100871, People's Republic of China}











\begin{abstract}
We have analyzed the broadband X-ray spectra of active galactic nuclei
(AGNs) in two ``non-merging'' luminous infrared galaxies (LIRGs)
UGC~2608 and NGC~5135, utilizing the data of \textit{NuSTAR},
\textit{Suzaku}, \textit{XMM-Newton}, and \textit{Chandra}.  Applying
the X-ray clumpy-torus model (XCLUMPY: \citealt{Tanimoto2019}), we find
that both sources have similar spectra characterized by 
Compton-thick (CT) absorption ($N_{\rm H} \sim$ 5--7 $\times
10^{24}$~cm$^{-2}$) and small torus angular width ($\sigma < 20\degr$).
The intrinsic 2--10~keV luminosities are $3.9^{+2.2}_{-1.7}$
$\times 10^{43}$ erg s$^{-1}$ (UGC~2608) and 
$2.0^{+3.3}_{-1.0}$ $\times 10^{43}$ erg s$^{-1}$ (NGC~5135). 
The [\ion{O}{4}]-to-nuclear-12~$\mu$m luminosity ratios are larger
than those of typical Seyferts, which are consistent with 
the torus covering factors ($C_{\rm T} \lesssim$ 0.7) 
estimated from the torus angular widths and column densities by
X-ray spectroscopy.
The torus covering factors and Eddington ratios ($\lambda_{\rm Edd} \sim$ 0.1)
follow the relation found by \citet{Ricci2017cNature} for local AGNs, 
implying that their tori become geometrically thin due to 
significant radiation pressure of the AGN that blows out some part of the tori.
These results indicate that the CT AGNs in these ``non-merger'' LIRGs are 
just a normal AGN population seen edge-on through 
a large line-of-sight column density.
They are 
in contrast to the buried CT AGNs in late-stage mergers that
have large torus covering factors even at large Eddington ratios.

\end{abstract}



\section{Introduction} 
\label{S_intro}

Luminous\footnote{LIRGs: $L_{\rm IR}$(8--1000 $\mu$m) 
$>$ 10$^{11}$\textit{L}$_{\odot}$ = 3.828 $\times 10^{44}$ erg s$^{-1}$} and 
ultraluminous\footnote{ULIRGs: $L_{\rm IR}$(8--1000 $\mu$m) 
$>$ 10$^{12}$\textit{L}$_{\odot}$ = 3.828 $\times 10^{45}$ erg s$^{-1}$} 
infrared galaxies (U/LIRGs; see \citealt{Sanders1996} for a review)
are key populations to understand the co-evolution of 
galactic bulges and their central supermassive black holes (SMBHs). 
They are powered by starburst and active galactic nucleus (AGN)
activities hidden behind gas and dust. The integrated infrared (IR) luminosity of U/LIRGs
constitutes a dominant fraction of the total radiation density
at $z>1$, which is much larger than that directly visible in the UV band
\citep{Madau2014}. Hence, they trace major
processes of star formation and SMBH growth in the cosmic history.

As a possible explanation for 
the cosmic ``downsizing'' of galaxies and SMBHs 
\citep{Cowie1996,Ueda2003}, it has been suggested that 
the triggering mechanisms of
star formation and mass accretion onto SMBHs 
may have two channels: more massive galaxies and SMBHs are formed 
via ``major mergers'' at higher redshifts, while less massive galaxies and 
SMBHs are former via ``secular processes'' at lower redshifts
\citep[e.g.,][]{Kormendy2011,Draper2012,Alexander2012,Barro2013,Tadaki2014}.
In fact, morphological studies revealed that LIRGs were generally a mix
of mergers and single isolated disk galaxies at $z<1$ \citep[e.g.,][]{Wang2006,Kartaltepe2010}. 
According to the above scenario, 
it is expected that the structure of AGNs in ``merging'' LIRGs and
``non-merging'' ones may be largely different because of the different
AGN triggering mechanisms, even if their IR luminosities (hence star
formation rates; SFRs) are similar. Previous studies of ``merging'' U/LIRGs
revealed that 
at late merger stages
the AGNs often became deeply ``buried'' 
with large torus covering factors 
by circumnuclear material of gas and dust (probed in the 1--80~keV 
broadband X-ray band; e.g., \citealt{Ricci2017bMNRAS}) 
and also of dust (probed in the 1--60~$\mu$m near-IR to far-IR band; 
e.g., \citealt{Sanders1988,Imanishi2008,Yamada2019}).
It is thus essential to investigate the properties of 
AGNs in ``non-merging'' LIRG utilizing the X-ray and IR data,
for a full understanding of whether or not
they have distinct characteristics from those in ``merging'' LIRGs.

In this work, we focus on AGNs in U/LIRGs 
with no signs of mergers from a complete flux-limited sample of the Great
Observatories All-sky LIRG Survey (GOALS; \citealt{Armus2009}), which
consists of 22 ULIRGs and 180 LIRGs having $f_{\nu}$(60 $\mu$m) 
$> 5.24$~Jy in the local ($z < 0.088$) universe.
The GOALS sample contains 12 non-merging LIRGs (according to
\citealt{Stierwalt2013}) that were observed with 
\textit{Nuclear Spectroscopic Telescope Array} 
(\textit{NuSTAR}; \citealt{Harrison2013}),
the first focusing telescope in orbit operating at $>$10~keV,
by Cycle-4. 
Among them, NGC~1068 and NGC~7130 are recently reported that they show features 
of past mergers
\citep{Davies2014,Tanaka2017}, and thus are excluded. We
also exclude NGC~1275 (a radio galaxy in the Perseus cluster) and NGC~1365 
(showing variable absorption; e.g., \citealt{Rivers2015}) from our sample as
exceptional sources. Performing the same data reduction in Section~\ref{Sub_nustar}, 
we find that the 6 objects, 
UGC~2612, NGC~4418, IC~860, NGC~5104, NGC~6907, and NGC~7591 
are either undetected or detected only below $\sim$10 keV with \textit{NuSTAR}.
Thus, we select UGC~2608 and NGC~5135
as the best targets for our study.

UGC~2608 ($z =$ 0.0233, log$L_{\rm
IR}/L_{\odot}$ = 11.4) and NGC~5135 ($z =$ 0.0137, log$L_{\rm
IR}/L_{\odot}$ = 11.3) are classified as nonmergers
from the \textit{Spitzer}/IRAC 3.6 $\mu$m and the high resolution
\textit{Hubble Space Telescope} (\textit{HST}) data \citep{Stierwalt2013}. 
Optically the sources were classified as Seyfert 2s \citep{Veron-Cetty2003}.
AGN signatures were also detected at the mid-IR wavelengths on the
basis of the 6.2 $\mu$m polycyclic aromatic hydrocarbon (PAH) equivalent
widths and the [\ion{Ne}{5}] 14.32~$\mu$m line luminosities \citep{Inami2013}.
The properties of AGNs and starbursts are constrained 
in the IR band by spectral energy distribution (SED) 
fitting \citep[e.g.,][]{Shangguan2019}.

The nature of the AGNs in UGC~2608 and NGC~5135 are still unclear
because of their heavily obscuration. Using \textit{XMM-Newton} data,
\citet{Guainazzi2005} suggested that UGC~2608 contains a 
heavily obscured AGN with 
$N_{\rm H} > 1.6 \times 10^{24}$~cm$^{-2}$, which is Compton-thick
(CT) level (i.e., $N_{\rm H} \gtrsim 1.5 \times 10^{24}$~cm$^{-2}$).
With \textit{Chandra},
\citet{Levenson2004} found that the AGN in NGC~5135 is heavily obscured
with $N_{\rm H} \geq 10^{24}$~cm$^{-2}$.
However, their intrinsic luminosities were not accurately estimated 
only from the X-ray observations below 10~keV.
\textit{Suzaku} detected narrow Fe K$\alpha$ lines 
of large equivalent widths (EWs), $0.6 \pm 0.3$~keV for UGC~2608 and
$1.6 \pm 0.4$~keV for NGC~5135, which are possible evidence for 
CT AGNs \citep{Fukazawa2011,Singh2012}, 
whereas the flux uncertainties above 10 keV are large due to the limited
photon statistics.
The more sensitive hard X-ray data of \textit{NuSTAR} observations enable us to characterize the properties of these AGNs.

In this paper, we analyze the best quality broadband X-ray spectra 
of UGC~2608 and NGC~5135, utilizing the data of \textit{NuSTAR}, 
\textit{Suzaku}, \textit{XMM-Newton}, and \textit{Chandra}. The
paper is organized as follows. In Section~\ref{S_reduction}, we describe
the observations and data reduction.  Section~\ref{S_analysis} reports
our results of the spectral analysis using two models: a baseline
model and a Monte Carlo-based model from a clumpy torus
(XCLUMPY; \citealt{Tanimoto2019}).  We discuss the implications from our
results in Section~\ref{S_discussion}.  Our findings are summarized in
Section~\ref{S_conclusion}.  Throughout the paper, we adopt the
cosmological parameters of $H_{\rm 0}$ = 70 km s$^{-1}$ Mpc$^{-1}$,
$\Omega_{\rm M}$ = 0.27, and $\Omega_{\rm \Lambda}$ = 0.73.  All errors
correspond to the 90\% confidence level for a single interesting
parameter.
\\

\begin{deluxetable*}{lllclc}
\tablewidth{\textwidth}
\tablecaption{Observation Log Utilized in This Work \label{T_log}}
\tablehead{
\colhead{Object}         &
\colhead{Satellite}      &
\colhead{Instrument}     &
\colhead{ObsID}          &
\colhead{UT Observation Date} &
\colhead{Net Exp. (ks)} 
}
\startdata
UGC~2608 &\textit{NuSTAR}     &FPMA, FPMB       &60001161002 &2014 Oct 08 17:06:07 & 22.1/22.2   \\
         &\textit{Suzaku}     &XIS-0, 1, 3      &701007020   &2007 Feb 04 18:21:52 & 39.5        \\
         &\textit{XMM-Newton} &EPIC/MOS1, 2, pn &0002942401  &2002 Jan 28 16:56:46 & 4.8/4.8/2.1 \\
NGC~5135 &\textit{NuSTAR}     &FPMA, FPMB       &60001153002 &2015 Jan 14 15:31:07 & 33.4/32.5   \\
         &\textit{Suzaku}     &XIS-0, 1, 3      &702005010	 &2007 Jul 03 05:59:41 & 52.5        \\
         &\textit{Chandra}    &ACIS-S           &2187        &2001 Sep 04 15:22:37 & 29.3
\enddata
\end{deluxetable*}

\section{Observations and Data Reduction}
\label{S_reduction}

The observation logs of UGC~2608 and NGC~5135 are 
summarized in Table~\ref{T_log}.
The X-ray spectral analysis is carried out by combining the 
\textit{NuSTAR} data (Section~\ref{Sub_nustar}) with those of \textit{Suzaku}
(Section~\ref{Sub_suzaku}), \textit{XMM-Newton} (Section~\ref{Sub_xmm}), 
and \textit{Chandra} (Section~\ref{Sub_chandra}).

\subsection{\textit{NuSTAR}}
\label{Sub_nustar}

\textit{NuSTAR} \citep{Harrison2013} is the first satellite capable of
focusing hard X-rays above 10 keV, covering the 3--79~keV band.  It
consists of two co-aligned X-ray telescopes coupled with focal plane modules
(FPMA and FPMB; FPMs).  \textit{NuSTAR} observed UGC~2608 for a net
exposure of $\sim$22 ks on 2014 October 8 (ObsID = 60001161002) and
NGC~5135 for $\sim$33 ks on 2015 January 14 (ObsID = 60001153002).  The
\textit{NuSTAR} data were processed with the \textit{NuSTAR} Data
Analysis Software \textsc{nustardas} v1.8.0 included in \textsc{heasoft}
v6.25, by adopting the calibration files released on 2019 May 13. The
\textsc{nupipeline} script was used to produce calibrated and cleaned
event files with the \textsc{saamode=optimized} and
\textsc{tentacle=yes} options.
The source spectra and light curves were extracted 
from a circular region with a radius of
50\arcsec\ by using the \textsc{nuproducts} task.
The background was taken from a nearby source-free circular
region with a radius of 100\arcsec.  
We confirmed that the spectra and light curves 
showed good agreement between FPMA and FPMB,
and then we coadded them to improve the photon statistics with the
\textsc{addascaspec} and \textsc{lcmath} tasks, respectively.  The
3--24~keV light curves of the sources show no evidence of significant flux
variability on a time scale longer than 5820 s (the orbital period 
of \textit{NuSTAR}). 
The spectral bins were grouped
to contain at least 50 and 60 counts for UGC~2608 and NGC~5135,
respectively, in order to facilitate the use of $\chi^2$ statistics
and not to lose energy resolution by overbinning.

\subsection{\textit{Suzaku}}
\label{Sub_suzaku}

UGC~2608 and NGC~5135 were observed with \textit{Suzaku}
\citep{Mitsuda2007} in 2007 February and July, respectively.
\textit{Suzaku}, the fifth Japanese X-ray satellite, carries four X-ray
CCD cameras called the X-ray Imaging Spectrometer (XIS-0, XIS-1, XIS-2,
and XIS-3), covering the energy band of
0.2--12~keV.  XIS-0, XIS-2, and XIS-3 are frontside-illuminated cameras
(XIS-FIs) and XIS-1 is a backside-illuminated one (XIS-BI).  XIS-2 data
were not available for both objects due to the malfunction that
had occurred in 2006
November. We used the \textit{Suzaku} calibration database released on
2018 October 23 for generating the cleaned event files
with the recommended filtering options. Source photons were extracted
from a circular region with a radius of 2.4\arcmin\ centered on the source
center, and the background was from two source-free circular regions
with radii of 2.2\arcmin.  
We generated the XIS response matrix files (RMF) with \textsc{xisrmfgen} and ancillary response files (ARF) with \textsc{xissimarfgen} \citep{Ishisaki2007}.
All the spectra of the XIS-FIs available in each
observation were combined together. The XIS spectra of UGC~2608 and
NGC~5135 were rebinned to contain at least 70 and 80 counts per energy
bin, respectively. 
We decided not to use the data of the Hard X-ray Detector (HXD)
in our analysis, due to the much poorer photon statistics 
than those of the \textit{NuSTAR} data in these faint targets and 
possible systematic uncertainties in the HXD background model \citep{Fukazawa2009}.

\subsection{\textit{XMM-Newton}}
\label{Sub_xmm}

The \textit{XMM-Newton} \citep{Jansen2001} observation of UGC~2608 was
performed on 2002 January 28.  We analyzed the data of two EPIC/MOS
(MOS1, MOS2) and EPIC/pn 
reprocessed with the \textit{XMM-Newton} Science Analysis System
(\textsc{sas}: \citealt{Gabriel2004}) v17.0.0 and Current Calibration
Files (CCF) of 2018 June 22. The raw MOS and pn data files were
reduced with the \textsc{emproc} and \textsc{epproc} tasks,
respectively. We selected good events with PATTERN $\leq$12 for MOS
and PATTERN $\leq$4 for pn. To filter out the periods of
background flares, we excluded data when the count rates
exceeded 1.0~counts~s$^{-1}$ in the 10--12~keV band for MOS and 
2.0~counts~s$^{-1}$ above 10~keV for pn. For both cameras, 
the source spectra were extracted from a circular region of 20\arcsec\ radius,
whereas the background ones were from a source-free region with a
radius of 40\arcsec\ in the same CCD chips. We created the RMF with 
\textsc{rmfgen} and ARF with \textsc{arfgen}. 
The source spectra, background spectra, RMFs, and ARFs
of EPIC/MOS1 and MOS2 were combined by using \textsc{addascaspec}, and
the spectral bins were merged to contain no less than 20 counts.

\subsection{\textit{Chandra}}
\label{Sub_chandra}

\textit{Chandra} \citep{Weisskopf2002} 
observed NGC~5135 with ACIS-S 
on 2001 September 04 for a net exposure of 29.3 ks.
The data reduction was performed with the standard procedures, by using
\textit{Chandra} Interactive Analysis of Observations (CIAO) v4.11 and
the Calibration Database (CALDB) v4.8.4.1. The event files were
reproduced by the \textsc{chandra\_repro} tool.  We extracted the source
spectrum from a circular region with a radius of 10.5\arcsec, and took the
background from a nearby source-free circular region with the same
radius. The spectrum was rebinned to have no less than 20 counts per bin.
\\

\begin{deluxetable*}{llllll}
\tablewidth{\textwidth}
\tablecaption{Summary of the Best-fit Spectral Parameters \label{T_parameters}}
\tablehead{
\colhead{No.} &
\colhead{Parameter} &
\colhead{\ \ \ \ \ \ UGC~2608} &
\colhead{ } &
\colhead{\ \ \ \ \ \ NGC~5135} &
\colhead{ } \\
 & &Baseline Model  &XCLUMPY Model &Baseline Model &XCLUMPY Model
}
\startdata
(1) &$N_{\rm H}^{\rm LOS}$ [$10^{24}$~cm$^{-2}$]	        &      $3.2_{-1.0}^{+0.4}$&      $5.4_{-3.1}^{+7.0}$&     $6.3_{-3.1}^{+3.7a}$&     $6.6_{-2.7}^{+22.5}$ \\
(2) &$N_{\rm H}^{\rm Equ}$ [$10^{24}$~cm$^{-2}$]        	&                  \nodata&          $15_{-9}^{+19}$&                  \nodata&     $9.5_{-3.9}^{+32.0}$ \\
(3) &$\Gamma_{\rm AGN}$                                 	&                  $1.8^b$&                  $1.8^b$&  $1.50_{-0.00}^{+0.06a}$&   $1.71_{-0.15}^{+0.19}$ \\
(4) &$A_{\rm AGN}$ [$10^{-2}$~keV$^{-1}$~cm$^{-2}$ s$^{-1}$]&   $0.65_{-0.42}^{+0.09}$&   $0.96_{-0.41}^{+0.54}$&   $0.59_{-0.49}^{+0.10}$&   $1.27_{-0.64}^{+2.06}$ \\
(5) &$f_{\rm scat}$ [\%]	                                &   $0.87_{-0.26}^{+1.62}$&   $0.57_{-0.24}^{+0.44}$&  $1.84_{-0.28}^{+8.16a}$&   $0.91_{-0.57}^{+0.77}$ \\
(6) &$\Gamma_{\rm scat}$                                	&  $3.00_{-0.43}^{+0.00a}$&  $2.87_{-0.59}^{+0.13a}$&   $2.79_{-0.10}^{+0.09}$&   $2.65_{-0.11}^{+0.15}$ \\
(7) &$R$	                                                &  $0.10_{-0.00}^{+0.18a}$&                  \nodata&  $0.10_{-0.00}^{+0.52a}$&                  \nodata \\
(8) &$\sigma$ [degree]	                                    &                  \nodata&    $10.0_{-0.0}^{+8.2a}$&                  \nodata&    $10.0_{-0.0}^{+4.5a}$ \\
(9) &$i$ [degree]	                                        &                   $60^b$&                   $80^b$&                   $60^b$&    $84.0_{-6.0}^{+3.0a}$ \\
(10) &$N_{\rm Line}$	                                    &                  \nodata&                  \nodata&                  \nodata&      $2.0_{-0.4}^{+0.5}$ \\
(11) &$k$T [keV]                                        	&   $0.82_{-0.06}^{+0.13}$&   $0.82_{-0.06}^{+0.14}$&   $0.90_{-0.03}^{+0.03}$&   $0.88_{-0.05}^{+0.03}$ \\
(12) &$A_{\rm apec}$ [10$^{-5}$ cm$^{-5}$]	                &   $4.58_{-0.84}^{+1.01}$&   $4.65_{-0.99}^{+1.06}$&   $6.22_{-0.79}^{+0.91}$&   $6.25_{-0.91}^{+1.06}$ \\
(13) &$C_{\rm FI}$	                                        &  $1.20_{-0.14}^{+0.00a}$&  $1.20_{-0.13}^{+0.00a}$&   $1.07_{-0.11}^{+0.12}$&   $1.03_{-0.13}^{+0.14}$ \\
(14) &$C_{\rm BI}$	                                        &  $1.20_{-0.11}^{+0.00a}$&  $1.20_{-0.11}^{+0.00a}$&  $1.12_{-0.13}^{+0.08a}$&  $1.08_{-0.14}^{+0.12a}$ \\
(15) &$C_{\rm pn}$	                                        &   $0.97_{-0.14}^{+0.15}$&   $0.97_{-0.15}^{+0.15}$&                  \nodata&                  \nodata \\
(16) &$C_{\rm ACIS}$                                    	&                  \nodata&                  \nodata&   $0.97_{-0.11}^{+0.11}$&   $0.94_{-0.12}^{+0.13}$ \\
(17) &$F_{2-10}^{\rm obs}$ [$10^{-13}$ erg s$^{-1}$~cm$^{-2}$] &                   2.1&                      2.1&                      3.8&                      3.7 \\
(18) &$F_{15-50}^{\rm obs}$ [$10^{-12}$ erg s$^{-1}$~cm$^{-2}$]&                   3.0&                      3.1&                      4.2&                      4.1 \\
(19) &$L_{\rm 2-10}$ [10$^{43}$ erg s$^{-1}$]               &      $3.2_{-2.0}^{+0.4}$&      $3.9_{-1.7}^{+2.2}$&      $1.6_{-1.3}^{+0.3}$&      $2.0_{-1.0}^{+3.3}$ \\
(20) &$L_{\rm 10-50}$ [10$^{43}$ erg s$^{-1}$]              &      $3.8_{-2.4}^{+0.5}$&      $5.0_{-2.2}^{+2.8}$&      $3.1_{-2.6}^{+0.5}$&      $3.0_{-1.5}^{+5.0}$ \\
(21) &$L_{\rm K\alpha}$ [10$^{40}$ erg s$^{-1}$]            &                      4.2&                      4.4&                      2.6&                      3.7 \\
(22) &$C_{\rm T}$                                           &                  \nodata&  $0.45_{-0.01}^{+0.31a}$&                  \nodata&  $0.44_{-0.01}^{+0.20a}$ \\
(23) &$C_{\rm T}^{(24)}$                                    &                  \nodata&  $0.28_{-0.03}^{+0.23a}$&                  \nodata&  $0.26_{-0.02}^{+0.21a}$ \\
&$\chi^2$/dof                                               &                  77.0/72&                  75.1/72&                 186.6/160&               164.1/158
\enddata
\tablecomments{Columns: 
(1) hydrogen column density along the line of sight;
(2) hydrogen column density along the equatorial plane (i.e., the maximum $N_{\rm H}$);
(3) power-law photon index of the AGN transmitted component;
(4) power-law normalization of the AGN transmitted component at 1~keV;
(5) scattering fraction;
(6) power-law photon index of the scattering component including emission lines from photoionized plasma;
(7) reflection strength ($R = \Omega/2\pi$) of the \textsf{pexmon} model;
(8) torus angular width;
(9) inclination angle of the torus;
(10) relative normalization of the emission lines from the torus;
(11) temperature of the \textsf{apec} model;
(12) normalization of the \textsf{apec} model;
(13) cross-calibration of \textit{Suzaku}/XIS-FIs relative to \textit{NuSTAR}/FPMs;
(14) cross-calibration of \textit{Suzaku}/XIS-BI relative to \textit{NuSTAR}/FPMs;
(15) cross-calibration of \textit{XMM-Newton} EPIC/pn relative to \textit{NuSTAR}/FPMs;
(16) cross-calibration of \textit{Chandra}/ACIS-S relative to \textit{NuSTAR}/FPMs;
(17--18) observed flux in the 2--10~keV and 15--50~keV band, respectively (normalized at the value of the \textit{NuSTAR} observation);
(19--20) intrinsic luminosity in the 2--10~keV and 10--50~keV band, respectively. 
The errors on luminosities are estimated by fixing the photon index at the best-fit value;
(21) luminosity of the iron-K$\alpha$ line; 
(22--23) torus covering factor derived from the equatorial column density 
and torus angular width through Equation (4) for Compton-thin ($N_{\rm H} \geq 10^{22}$ cm$^{-2}$) and Compton-thick ($N_{\rm H} \geq 10^{24}$ cm$^{-2}$) material, respectively.
}
\tablenotetext{a}{The parameter reaches a limit of its allowed range.}
\tablenotetext{b}{Value fixed.}
\end{deluxetable*}

\section{X-ray Spectral Analysis}
\label{S_analysis}

The broadband X-ray spectra of these sources ($\sim$0.5--70~keV) 
consisting of multiple instrument data are
simultaneously analyzed. Considering the signal-to-noise ratio of the spectra, 
we determine the energy band to be used as follows: for
UGC~2608, \textit{NuSTAR}/FPMs (3--70~keV), \textit{Suzaku}/XIS-FIs
(0.9--7~keV), XIS-BI (0.45--7~keV), \textit{XMM-Newton} EPIC/MOS
(0.7--7~keV), and EPIC/pn (0.5--7~keV); for NGC~5135,
\textit{NuSTAR}/FPMs (3--74~keV), \textit{Suzaku}/XIS-FIs (0.8--10~keV),
XIS-BI (0.7--8~keV), and \textit{Chandra}/ACIS-S (0.5--6~keV). The data
of XIS-FIs and XIS-BI in the 1.6--1.9~keV band are excluded to avoid
the calibration uncertainty.\footnote{https://heasarc.gsfc.nasa.gov/docs/suzaku/analysis/abc/node8.html} 
The observed spectra folded with the energy
responses are plotted in the left panels of Figure~\ref{F_UGC2608} and
\ref{F_NGC5135}.

In spectral fitting, we consider the Galactic absorption by multiplying
\textsf{phabs} to intrinsic spectral models, and fix the hydrogen column
densities at 2.03 $\times 10^{21}$~cm$^{-2}$ for UGC~2608 and 6.02
$\times 10^{20}$~cm$^{-2}$ for NGC~5135 \citep{Willingale2013}. We correct
possible cross-calibration uncertainties among different instruments by
introducing constant factors (\textsf{const1}; $C_{\rm FI}$, $C_{\rm
BI}$, $C_{\rm pn}$, and $C_{\rm ACIS}$ for XIS-FIs, XIS-BI, EPIC/pn, and
ACIS-S, respectively).  This value is set to unity for the
\textit{NuSTAR}/FPMs and \textit{XMM-Newton} EPIC/MOS as calibration
references, both of which are well cross-calibrated within a few percent
\citep{Madsen2015,Madsen2017}, and allowed the other constants to vary
within a range of 0.8--1.2. The solar abundances by
\citet{Anders1989} are assumed. Possible time variability of the AGN
emission is ignored because it is not significantly required from the
data. All spectral fitting was carried out on \textsc{xspec} v12.10.1
\citep{Arnaud1996} by adopting $\chi^2$ statistics.

\subsection{Baseline Model}

Previous X-ray studies suggested that UGC~2608 and NGC~5135 hosted CT AGNs
\citep[e.g.,][]{Guainazzi2005,Fukazawa2011,Singh2012}, and therefore we
start with a conventional model often applied for CT objects
\citep[e.g.,][]{Oda2017,Ricci2017dApJS,Tanimoto2018}.  The model
consists of four components: an absorbed direct component, an
unabsorbed scattered component, a reflection component, and a soft
thermal component. In the \textsc{xspec} terminology, it is described as
\begin{align}
&\mathsf{const1 * phabs * (zphabs * cabs * zpowerlw * zhighect} \notag\\
&\mathsf{+ const2 * zpowerlw * zhighect + pexmon + apec)}. 
\end{align}
The first term represents the transmitted AGN component modeled by an 
absorbed power law with a high-energy cutoff of 300~keV 
\citep[e.g.,][]{Dadina2008}.  We consider the Compton scattering effect 
to the primary component by multiplying the \textsf{cabs} model. 
For UGC~2608, the photon index ($\Gamma_{\rm AGN}$) is fixed at 1.8 as a 
typical value \citep[e.g.,][]{Ueda2014,Ricci2017dApJS}, 
since it cannot be well constrained due to the limited photon statistics.
The second term is the component representing scattering by ionized material. 
The normalization and cutoff energy are linked to those of the primary
component. The photon index ($\Gamma_{\rm scat}$) 
is set to vary between 1.5 and 3.0, because photoionized plasma
often contains emission lines that cannot be resolved by CCD, making the
apparent slope steeper than that of the primary component.
We multiply the scattering fraction (\textsf{const2}, $f_{\rm scat}$)
relative to the transmitted emission at 1 keV \citep[e.g.,][]{Ueda2007}.
The third term describes the reflection component from the torus. 
For Compton-reflection continuum we use the \textsf{pexmon} model in
XSPEC \citep{Nandra2007}, which calculates a reflection spectrum from
optically-thick cold matter (\textsf{pexrav};
\citealt{Magdziarz1995}) including Fe K$\alpha$ (6.4~keV), Fe K$\beta$ 
(7.1~keV), and Ni K$\alpha$ (7.5~keV) fluorescence lines. 
The relative intensity, $R$ = $\Omega$/2$\pi$ ($\Omega$ is the solid
angle of the reflector), is allowed to vary 
within a range of $0.1 \leq R \leq 2$ in order to avoid unrealistic
solutions, 
and the inclination angle is set to $i$ = 60\degr.
The normalization, photon index, and cutoff energy are tied to 
those of the primary component.
The last term accounts for optically-thin thermal emission from plasma 
in the host galaxy (\textsf{apec}; \citealt{Smith2001}), likely 
related to star-forming regions.

This analytic model well reproduces the broadband spectra of UGC~2608
(with the chi-squared statistic divided by the number of degrees of freedom, $\chi^2$/dof = 77.0/72) and NGC~5135 ($\chi^2$/dof = 186.6/160) in the
$\sim$0.5--70~keV energy band. The best-fit parameters 
are summarized in Table~\ref{T_parameters}. From the best-fit
model, we calculate the observed fluxes, intrinsic AGN luminosities, and Fe
K$\alpha$ luminosities, which are also listed in Table~\ref{T_parameters}.
The line-of-sight column densities are estimated to be $N_{\rm H}^{\rm
LOS}$ = $3.2^{+0.4}_{-1.0} \times 10^{24}$~cm$^{-2}$ (UGC 2608) and
$>$3.2 $\times 10^{24}$~cm$^{-2}$ (NGC 5135), classifying both 
objects as CT AGNs. The folded spectra and best-fit models are plotted in
the upper panels of Figure~\ref{F_UGC2608} and \ref{F_NGC5135}.

\subsection{XCLUMPY Model}

Next, we apply the XCLUMPY model \citep{Tanimoto2019}, a Monte 
Carlo-based spectral model from the clumpy torus in an AGN. Many
observations in the IR band indicate that AGN tori must be 
composed of dusty clumps rather than of a
smooth mixture of gas and dust
\citep[e.g.,][]{Krolik1988,Wada2002,Honig2007}.
Moreover, X-ray works using smooth torus models often suggest
a large amount of unabsorbed reflection 
components noticeable below the Fe K edge.
These spectral
features are interpreted to be caused by clumpy tori
\citep[e.g.,][]{Liu2014,Furui2016,Tanimoto2018,Tanimoto2019}. 
Therefore, we regard it as
one of the most physically realistic models currently available.

The biggest advantage of the XCLUMPY model is that the same
model can be fitted to the X-ray and the IR data. In this model,
the geometry of the torus is assumed to be the same as that in 
the CLUMPY model in the IR band
\citep{Nenkova2008a,Nenkova2008b}, where clumps are distributed
according to power-law and normal distributions in the radial and
angular directions, respectively.
In reality, the gas and dust are not 
located in the same place.\footnote{\citet{Ichikawa2019} argued
a possible presence of X-ray absorbing dust-free gas.
\citet{Tanimoto2020} suggested that the IR spectrum of an AGN is largely affected by a dusty polar outflow.}
Then, a comparison between self-consistent X-ray and IR models provides more robust measures of the covering factors for the ``gas'' (probed in the X-ray band) and the ``dust'' (probed in the IR band).

The geometry in the XCLUMPY model is determined by the column
density along the equatorial plane ($N_{\rm H}^{\rm Equ}$), the torus
angular width ($\sigma$ within a range of $10^{\circ}$--$70^{\circ}$),
and the inclination angle ($i$ within a range of
$20^{\circ}$--$87^{\circ}$).\footnote{The other fixed parameters are
the inner and outer radii of the torus (0.05 pc and 1.00 pc), the radius
of each clump (0.002 pc), the number of clumps along the equatorial plane
(10.0), and the index of radial density profile (0.5)
\citep[see][]{Tanimoto2019}.} In \textsc{XSPEC}, the model is
written as:
\begin{align}
&\mathsf{const1 * phabs * (zphabs * cabs * zpowerlw * zhighect} \notag\\
&\mathsf{+ const2 * zpowerlw * zhighect + atable\{xclumpy\_R.fits\}} \notag\\
&\mathsf{+ const3 * atable\{xclumpy\_L.fits\} + apec)}. 
\end{align}
The first (transmitted component) and second (scattered component by ionized material) 
terms are the same as in the previous model.
The third and fourth ones represent the two table models of XCLUMPY, 
the reflection continuum (xclumpy\_R.fits) and fluorescence lines 
(xclumpy\_L.fits), respectively.
The parameters of the reflection continuum and lines are fixed to be the same.
The relative normalization of the emission lines to the reflection
continuum ($N_{\rm Line}$; \textsf{const3}) is set to be a free
parameter for NGC~5135, which 
significantly improves the fit ($\Delta\chi^2 =$ 30.4 for 1 dof).
This factor can take into account uncertainties caused by the
assumption of geometry and possible non-solar abundances.
The photon index, cutoff energy, and normalization of the power law are
linked to those of the transmitted component. The inclination of
UGC~2608 is fixed at 80\degr, which cannot be well determined with 
the data. The line-of-sight column
density ($N_{\rm H}^{\rm LOS}$) in the first term is linked to 
the equatorial column density ($N_{\rm H}^{\rm Equ}$), the inclination, 
and the torus angular width through Equation
(3) in \citet{Tanimoto2019}. The fifth term corresponds to the emission
from optically thin thermal plasma in the host galaxy.

\begin{figure*}
    \epsscale{1.105}
    \plottwo{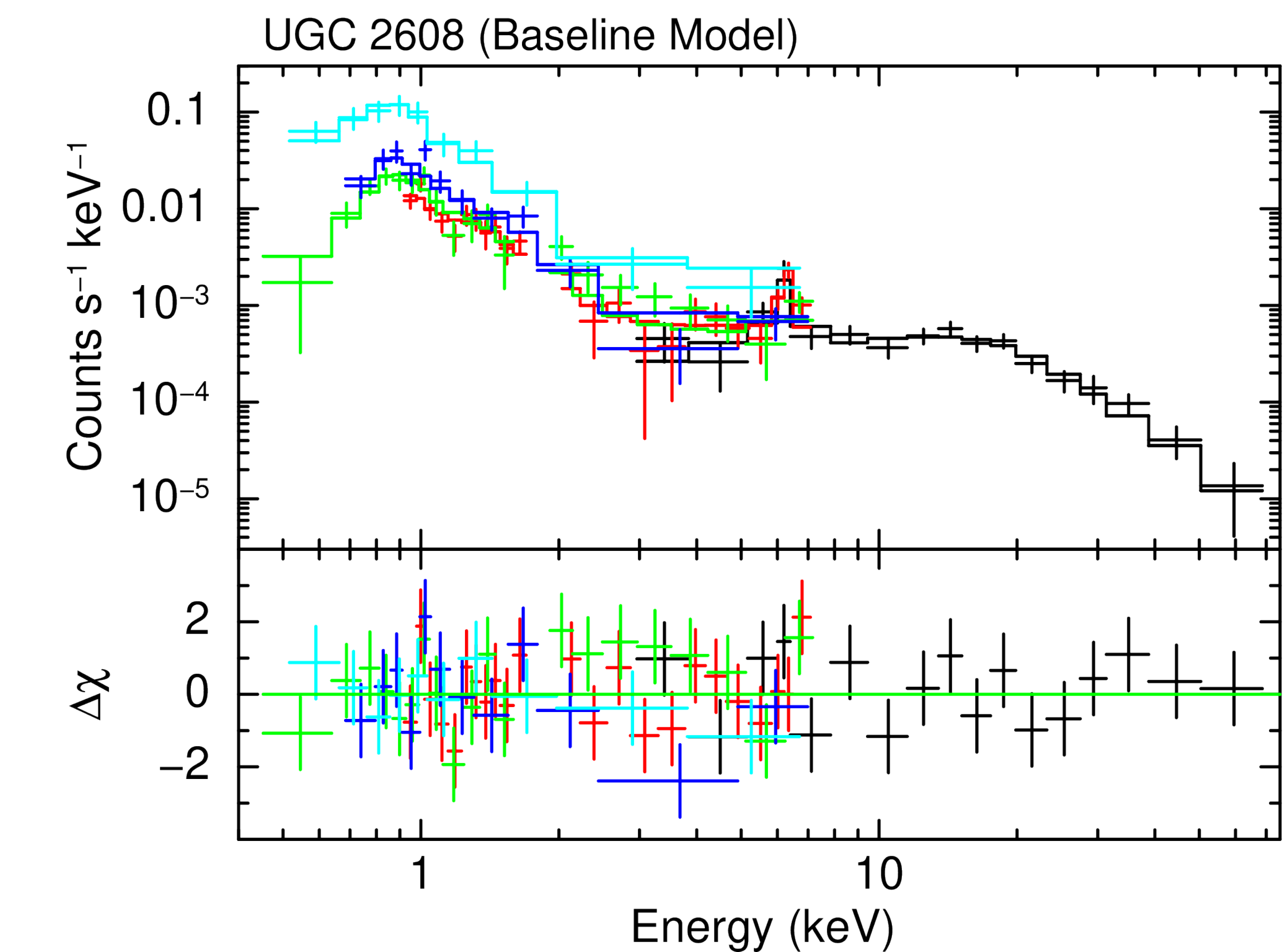}{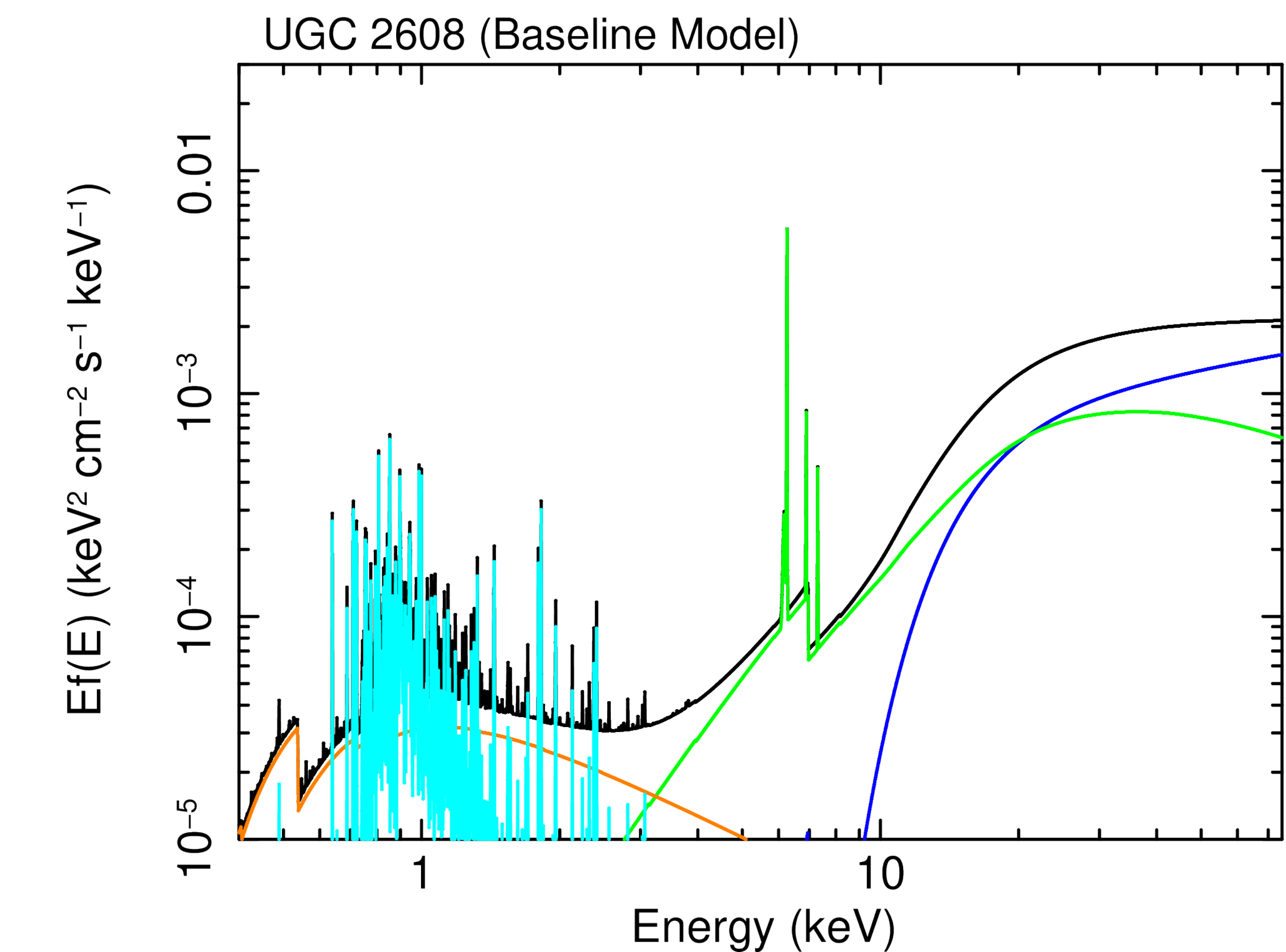}
    \plottwo{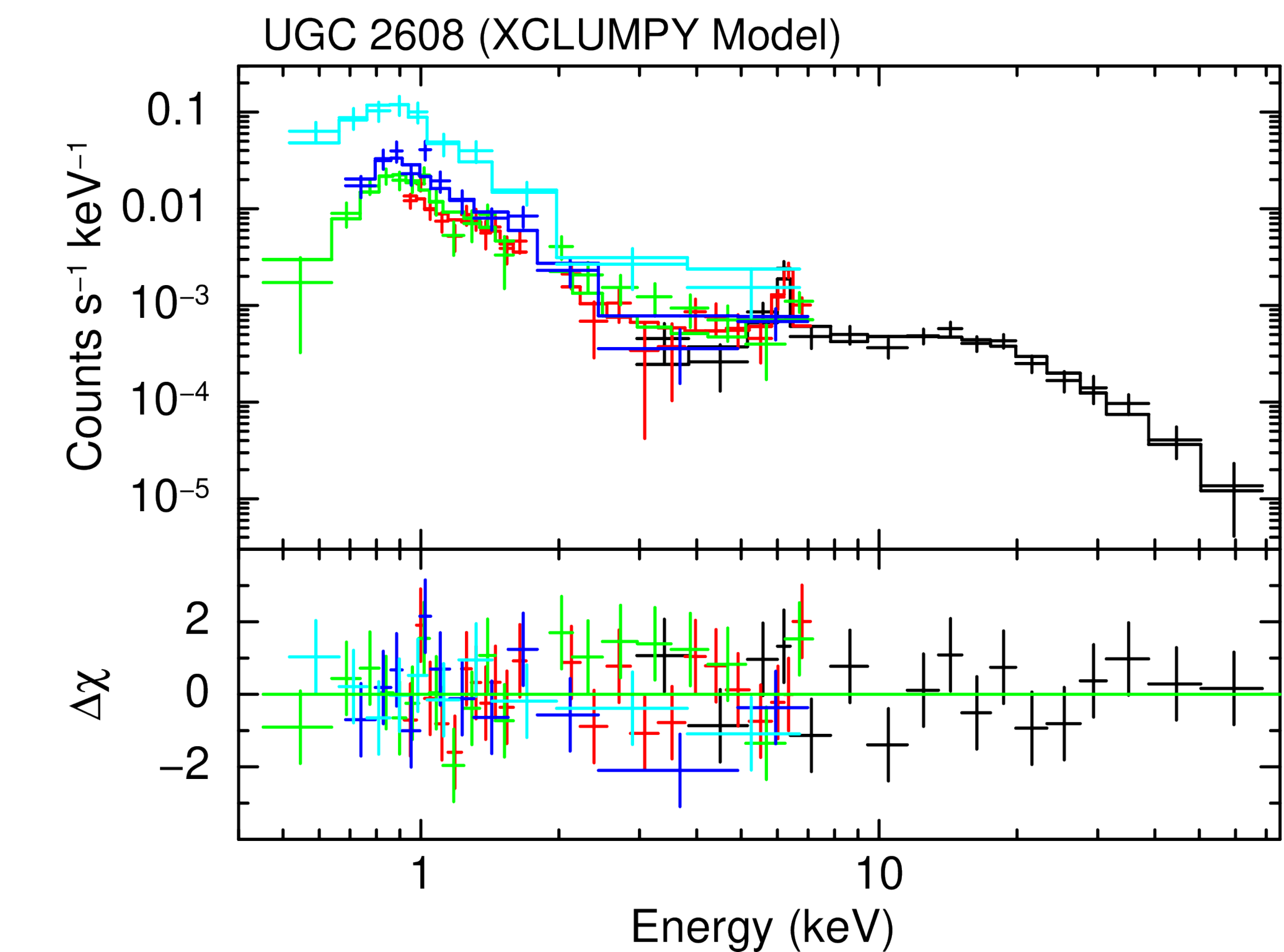}{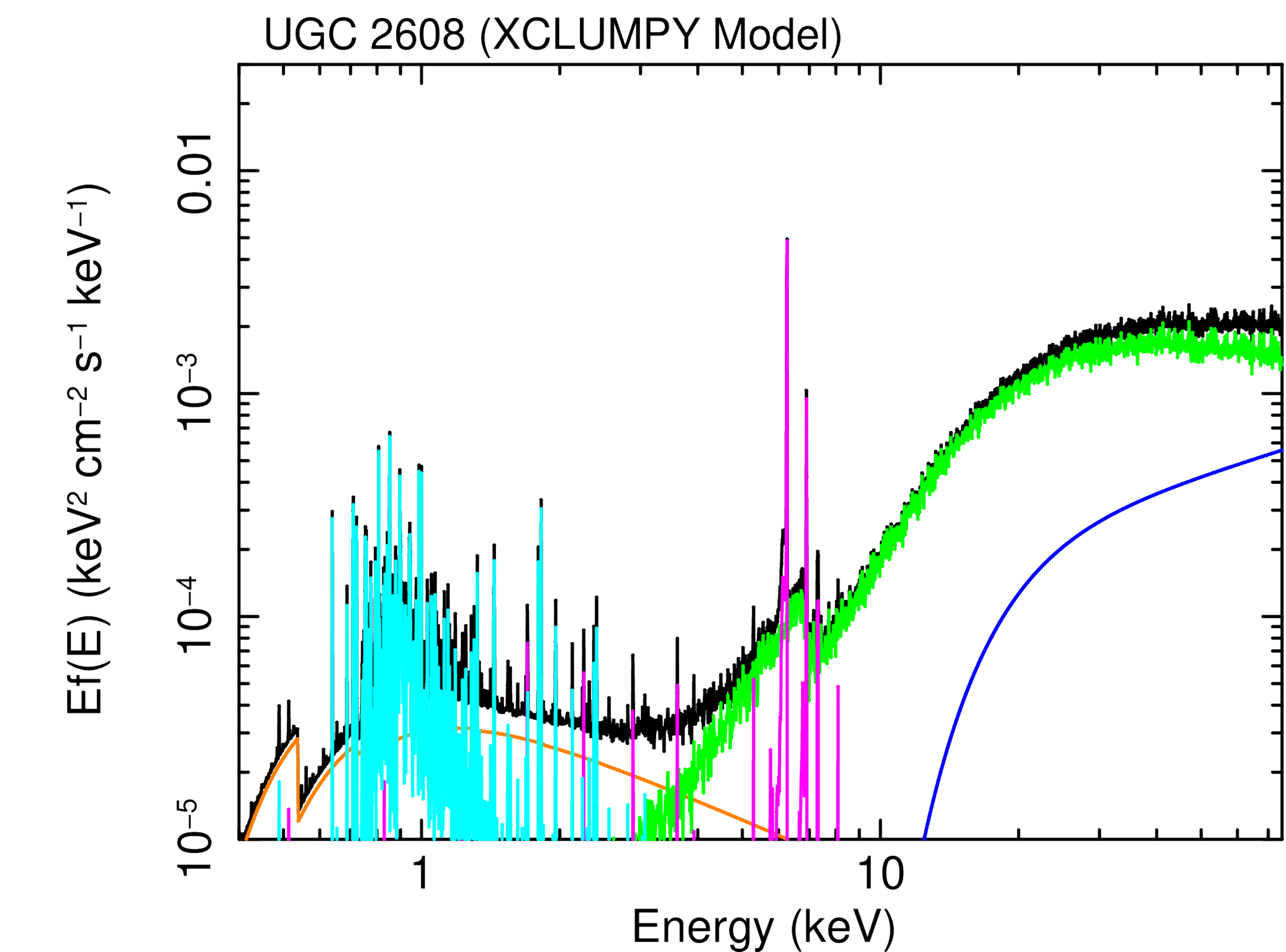}
    \caption{Folded spectra and best-fit models of UGC~2608 with the baseline model (upper panels) and XCLUMPY model (lower panels). Left panels: the black, red, green, blue, and cyan crosses are the data of \textit{NuSTAR}/FPMs, \textit{Suzaku}/XIS-FIs, XIS-BI, \textit{XMM-Newton} EPIC/MOS, and EPIC/pn, respectively.
    The solid lines represent the best-fit models, and the bottom panels show the residuals.
    Right panels: the black, blue, orange, green, magenta, cyan lines
 represent the total, AGN transmitted component, scattering component,
 reflection component, emission lines from the torus, and thermal
 emission from the host galaxy, respectively, 
in units of $EI_{E}$.\\
    }
    \label{F_UGC2608}
\end{figure*}

\begin{figure*}
    \epsscale{1.105}
    \plottwo{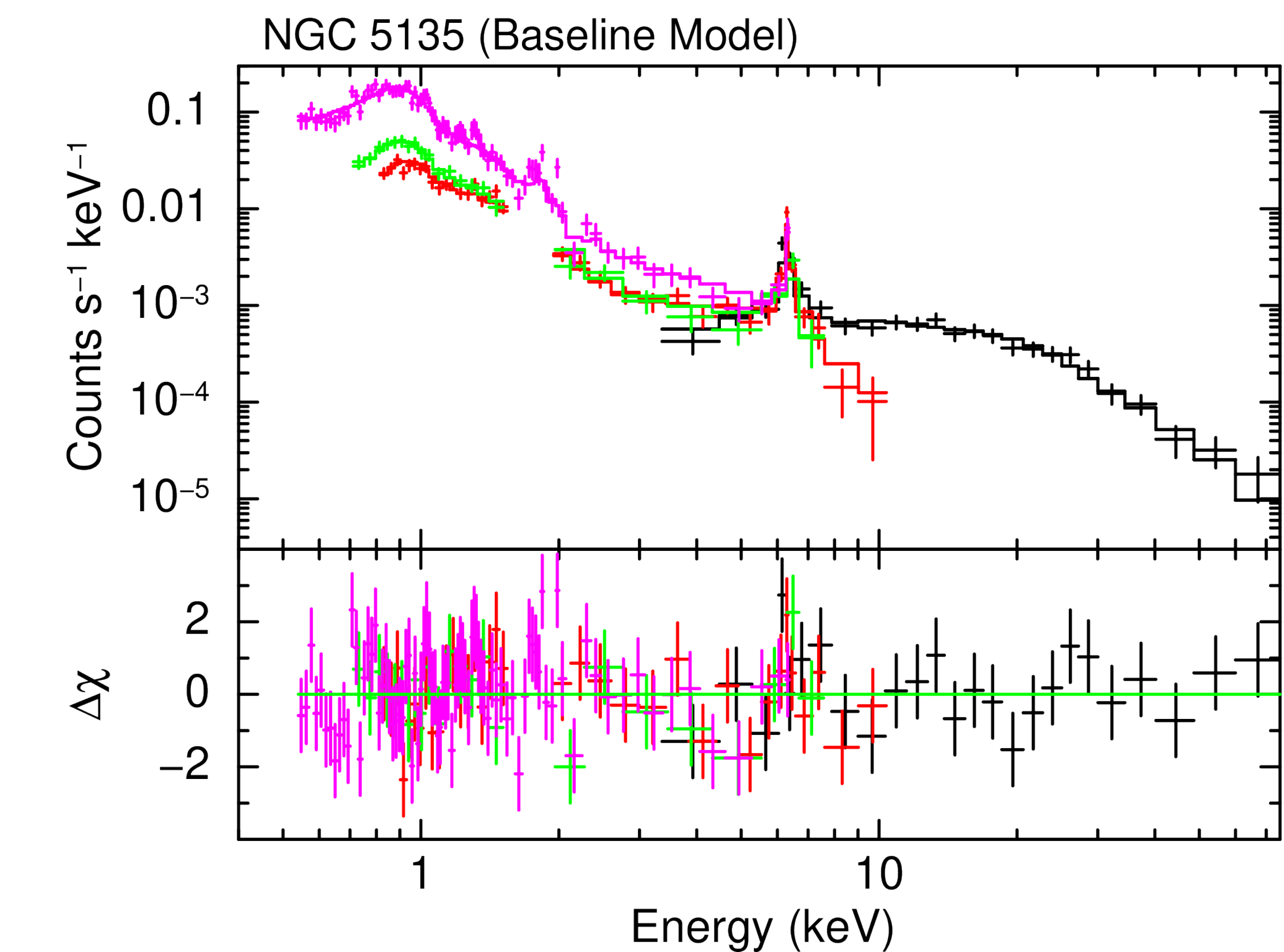}{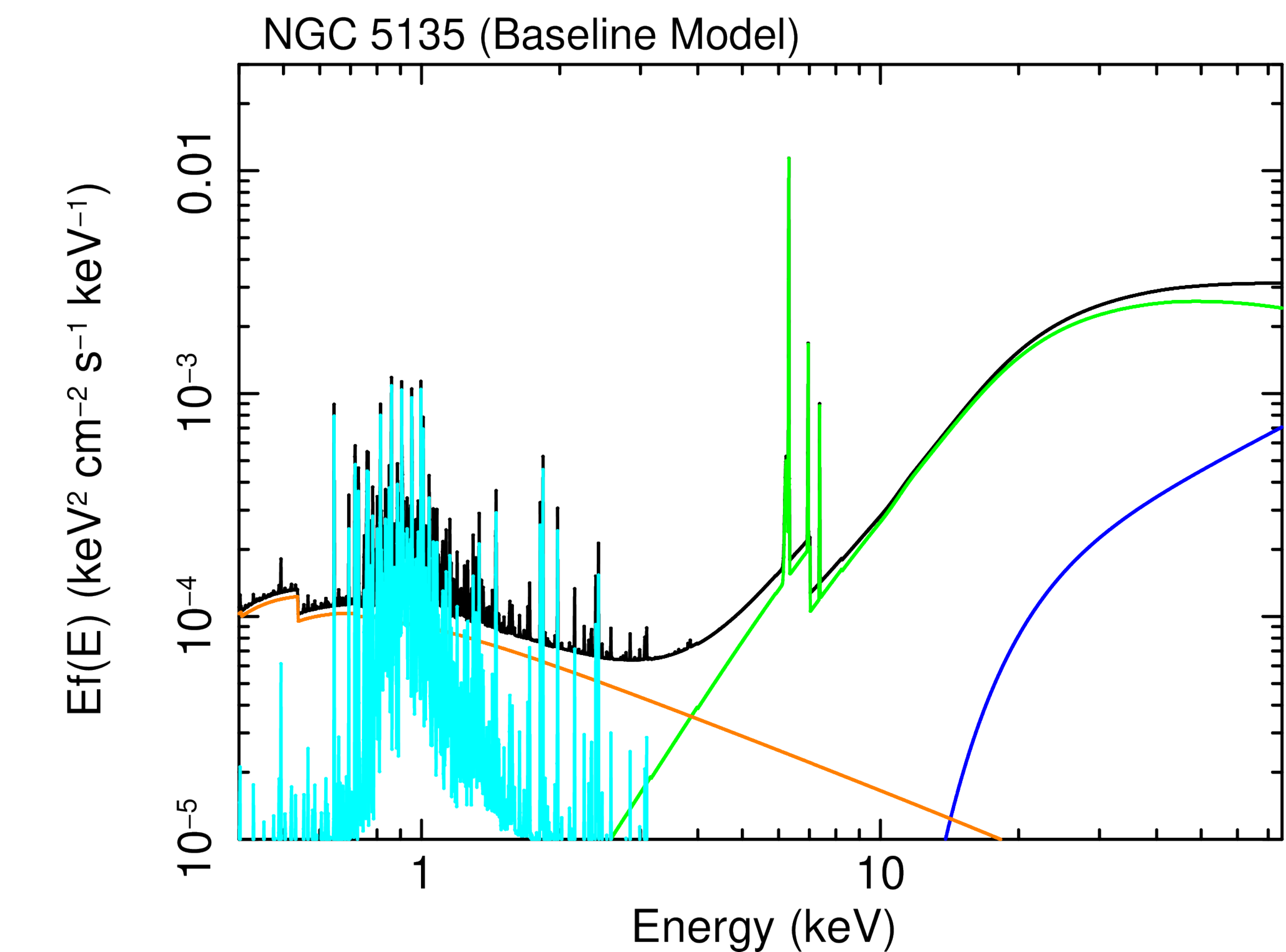}
    \plottwo{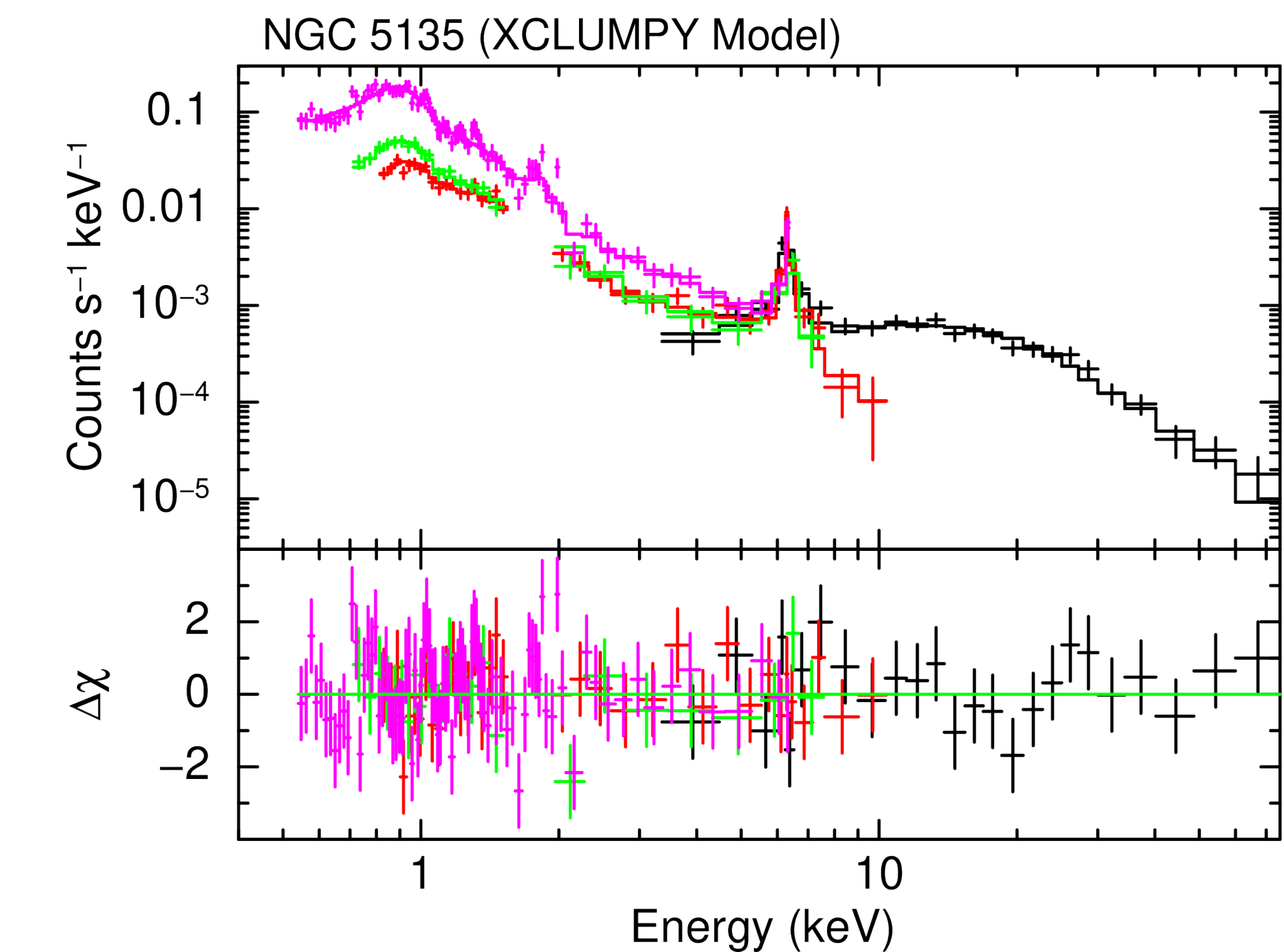}{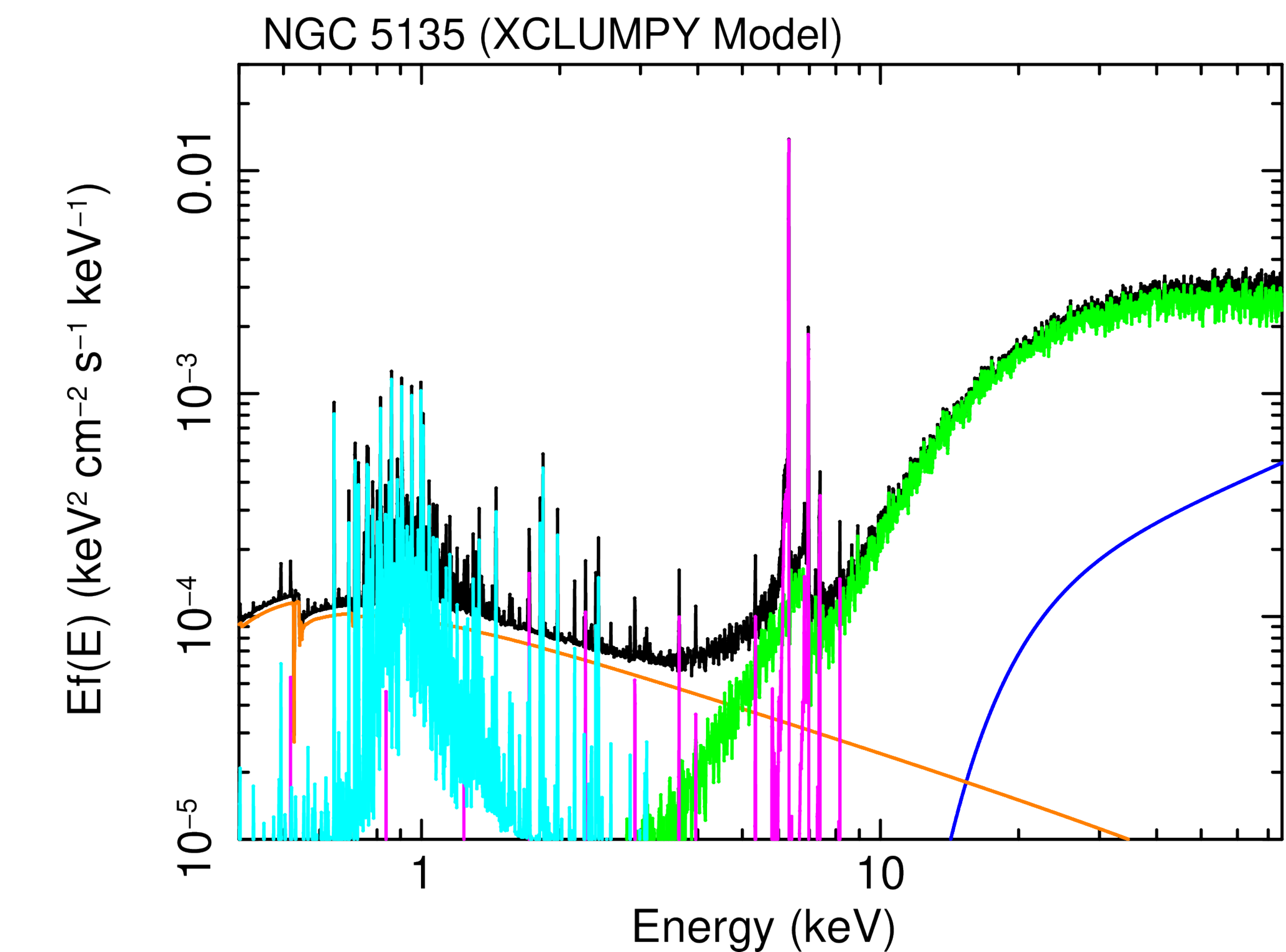}
    \caption{Folded spectra and best-fit models of NGC~5135 with the baseline model (upper panels) and XCLUMPY model (lower panels). Left panels: the black, red, green, and magenta crosses are the data of \textit{NuSTAR}/FPMs, \textit{Suzaku}/XIS-FIs, XIS-BI, and \textit{Chandra}/ASIS-S, respectively.
    The solid lines represent the best-fit models, and the bottom panels show the residuals.
    Right panels: the black, blue, orange, green, magenta, cyan lines
 represent the total, AGN transmitted component, scattering component,
 reflection component, emission lines from the torus, and thermal
 emission from the host galaxy, respectively, 
in units of $EI_{E}$.\\
    }
    \label{F_NGC5135}
\end{figure*}

We find this model also well reproduces the spectra of UGC~2608
($\chi^2$/dof = 75.1/72) and NGC~5135 ($\chi^2$/dof = 164.1/158). These
fits are statistically better compared with those by the baseline 
model, although both models represent statistically 
acceptable representations of the data. Table~\ref{T_parameters} lists
the best-fit parameters, observed fluxes, intrinsic X-ray luminosities, and
K$\alpha$ luminosities. The lower panels of Figure~\ref{F_UGC2608} and
\ref{F_NGC5135} illustrate the folded spectra and best-fit models.  The
line-of-sight column densities are found to be
$N_{\rm H}^{\rm LOS}$ = $5.4^{+7.0}_{-3.1} \times 10^{24}$~cm$^{-2}$
and $6.6^{+22.5}_{-2.7} \times 10^{24}$~cm$^{-2}$,
and the best-fit intrinsic 2--10~keV luminosities are 
$L_{\rm 2-10} = 3.9^{+2.2}_{-1.7} \times 10^{43}$~erg~s$^{-1}$ and
$2.0^{+3.3}_{-1.0} \times 10^{43}$~erg~s$^{-1}$ for UGC~2608 and
NGC~5135, respectively. 
This confirms that both sources host CT AGNs.
\\

\section{Discussion}
\label{S_discussion}

We have presented the best quality broadband X-ray spectra of
UGC~2608 and NGC~5135 in the $\sim$0.5--70~keV band, observed with
\textit{NuSTAR}, \textit{Suzaku}, \textit{XMM-Newton} and
\textit{Chandra}.  The sensitive hard X-ray data of \textit{NuSTAR} have
enabled us to best constrain the properties of the obscured AGNs.
For both targets, the 
XCLUMPY model better reproduces 
the combined spectra than the baseline model, 
which are found to be dominated by
the reflection component from the torus above 10~keV. 
The resultant line-of-sight column densities and intrinsic 2--10~keV luminosities
are consistent between the two models. Hereafter, we refer to
the results obtained with the XCLUMPY model, where a more realistic geometry
is considered.

We have revealed that UGC~2608 and NGC~5135 contain heavily CT AGNs with
column densities of $N_{\rm H}^{\rm LOS}$ = $5.4^{+7.0}_{-3.1}
\times 10^{24}$~cm$^{-2}$ and $6.6^{+22.5}_{-2.7} \times
10^{24}$~cm$^{-2}$, respectively. 
The large Fe K$\alpha$ EWs,
$\sim$1.4~keV for UGC~2608 and $\sim$2.5~keV for NGC~5135
(see also \citealt{Fukazawa2011}), 
are characteristics of CT AGNs \citep{Ghisellini1994,Ricci2014}.
Our results thus confirm the main conclusions of previous X-ray studies.
Using the
\textit{XMM-Newton} spectra of UGC~2608, \citet{Guainazzi2005} obtained
a photon index of $\Gamma = 2.3^{+1.1}_{-0.3}$, which is assumed to be
common for the AGN and scattered components, and a hydrogen column
density of $N_{\rm H} > 1.6 \times 10^{24}$~cm$^{-2}$.
Analyzing the \textit{Suzaku} 0.5--50~keV spectra 
of NGC~5135, \citet{Singh2012} obtained 
$\Gamma_{\rm AGN} = 1.65^{+0.26}_{-0.12}$ and 
$\Gamma_{\rm scat} = 2.31^{+0.12}_{-0.07}$ with a column density of
$N_{\rm H} = 2.20^{+0.36}_{-0.27} \times 10^{24}$~cm$^{-2}$. This column density of
NGC~5135 is smaller by a factor of $\sim$3 than our result. 
We infer that the difference may be due to a calibration issue and/or 
time variability in $N_{\rm H}$, 
since the \textit{Suzaku} HXD/PIN flux in the 15--50~keV 
band\footnote{The error includes a systematic uncertainty of
$\lesssim$3\% in the HXD/PIN background model \citep{Fukazawa2009}. 
The best-fit photon index of 1.71 is assumed to
convert from the observed count rate into the 15--50 keV flux.}, 
$F_{\rm 15-50}^{\rm obs} = (1.09 \pm 0.13) \times 10^{-11}$~erg~cm$^{-2}$~s$^{-1}$, was $\sim$3
times larger than the best-fit \textit{NuSTAR} flux.

\subsection{Luminosities and Eddington Ratios}
\label{Sub_bh}

In the IR band, the presence of the obscured AGNs in UGC~2608
and NGC~5135 was suggested by detections of high-excitation
emission lines, in particular of [\ion{O}{4}] 25.89 $\mu$m and [\ion{Ne}{5}]
14.32~$\mu$m \citep{Inami2013}; using the five diagnostics of
\textit{Spitzer}/IRS data\footnote{The diagnostics are following: the
[\ion{Ne}{5}]$_{14.3}$/[\ion{Ne}{2}]$_{12.8}$ flux ratios; the
[\ion{O}{4}]$_{25.9}$/[\ion{Ne}{2}]$_{12.8}$ flux ratios; the equivalent
width of the 6.2 $\mu$m PAH; the $S_{30}$/$S_{15}$ dust continuum slope;
and the Laurent diagram \citep{Laurent2000}.}, 
\citet{Diaz-Santos2017} estimated
the AGN fractions in the mid-IR luminosities
to be $\sim$62\% for UGC~2608 and
$\sim$36\% for NGC~5135 (the average of the five methods). 
Spectral decomposition can constrain the physical 
properties of the AGN in more detail. 
From the broadband IR ($\sim$1--500~$\mu$m) SED fitting, 
\citet{Shangguan2019} estimated 
the stellar masses, log$(M_{\rm stellar}/M_{\odot}) = 10.92 \pm 0.20$ 
and $11.03 \pm 0.20$, the SFRs, 
log(SFR/$M_{\odot}$yr$^{-1}$) = $1.37 \pm 0.20$ 
and $1.34 \pm 0.01$, and the IR luminosities from the tori, 
log($L_{\rm torus}^{\rm (IR)}$/erg s$^{-1}$) = $44.18^{+0.05}_{-0.04}$
and $43.76^{+0.03}_{-0.03}$ for UGC~2608 and NGC~5135, respectively. 
The intrinsic 2--10~keV luminosity of a normal Seyfert galaxy 
can be estimated from an IR AGN luminosity 
by using the relation by \citet{Mullaney2011}:
\begin{align}
&{\rm log}L_{\rm 2-10,43} = \frac{{\rm log}L_{\rm IR,43}^{\rm AGN} -
 (0.53 \pm 0.26)}{(1.11 \pm 0.07)},
\end{align}
where the X-ray and IR luminosities are in units of $10^{43}$~erg~s$^{-1}$.
Assuming $L_{\rm torus}^{\rm (IR)} \approx L_{\rm IR}^{\rm AGN}$, 
the above equation yields 
X-ray luminosities of $L_{\rm 2-10}$ = $(3.9 \pm
2.6) \times 10^{43}$ erg s$^{-1}$ (UGC~2608) and $(1.6 \pm 1.1) \times
10^{43}$~erg~s$^{-1}$ (NGC~5135). They are in good agreement with 
our results derived from the broadband X-ray spectral analysis.

The BH masses ($M_{\rm BH}$) of UGC~2608 and NGC~5135 are 
estimated as log($M_{\rm BH}/M_{\odot}$) $\sim 7.78 \pm 0.50$
\citep{Dasyra2011} and $\sim 7.29 \pm 0.44$ \citep{Marinucci2012},
respectively, on the basis of the $M_{\rm BH}$-$\sigma_{\*}$ relation.\footnote{
These BH masses are consistent with those calculated 
from the stellar masses of the host galaxies
by using the empirical relation of \citet{Reines2015}.}
Adopting a 2--10 keV-to-bolometric luminosity ratio of 20, 
a typical value for Seyferts \citep{Vasudevan2007}, 
we can convert our best-fit X-ray luminosities to 
the bolometric ones as $L_{\rm bol}^{\rm AGN} =$ 
$7.8^{+4.4}_{-3.3}$ $\times 10^{44}$ erg s$^{-1}$ (UGC~2608) 
and $4.1^{+6.6}_{-2.1}$ $\times 10^{44}$ erg s$^{-1}$ (NGC~5135).
Then, the Eddington ratios are calculated to be 
$\lambda_{\rm Edd} =$ $0.10^{+0.25}_{-0.07}$ (UGC~2608) and
$0.17^{+0.51}_{-0.12}$ (NGC~5135). 

\begin{deluxetable}{lcclc}
\tablewidth{\textwidth}
\tablecaption{Nuclear 12 $\mu$m and Intrinsic 2--10~keV Luminosity for AGNs in the Local U/LIRGs \label{T_12um}}
\tablehead{
\colhead{Object\ \ \ \ \ \ \ \ \ \ \ \ \ \ } &
\colhead{M} &
\colhead{log$L_{\rm 12\,\mu m}^{\rm (nuc)}$} &
\colhead{log$L_{\rm 2-10}$} &
\colhead{Ref} \\
(1)&(2)&(3)&\ \ \ \ \ \ (4)&(5)
}
\startdata
NGC 34          &D &$43.08 \pm 0.05$   &$41.98 \pm 0.60$ &1 \\
NGC 235A        &B &$43.31 \pm 0.20$   &$43.18 \pm 0.30$ &1 \\
NGC 1068        &N &$43.80 \pm 0.15$   &$43.64 \pm 0.30$ &1 \\
NGC 1365        &N &$42.54 \pm 0.04$   &$42.12 \pm 0.20$ &1 \\
IRAS 05189--2524&D &$44.87 \pm 0.17$   &$43.73 \pm 0.43$ &1 \\
UGC 5101        &D &$44.35 \pm 0.08$   &$43.15 \pm 0.40^{a}$ &2 \\
NGC 3690W       &C &$43.73 \pm 0.28$   &$43.26 \pm 0.60$ &1 \\
NGC 3690E       &C &$43.30 \pm 0.24$   &$39.76 \pm 0.30$ &1 \\
IC 883          &D &$<$43.90           &$40.99 \pm 0.40^{a}$ &3 \\
MCG-03-34-064   &A &$44.00 \pm 0.05$   &$43.33 \pm 0.48$ &1 \\
IC 4518W        &B &$43.54 \pm 0.07$   &$43.05 \pm 0.59$ &1 \\
ESO 286--19     &D &$44.67 \pm 0.04$   &$42.39 \pm 0.30$ &1 \\
NGC 7130        &N &$43.17 \pm 0.09$   &$42.81 \pm 0.60$ &1 \\
NGC 7469        &A &$43.83 \pm 0.05$   &$43.19 \pm 0.07$ &1 \\
NGC 7674        &A &$44.26 \pm 0.06$   &$44.02 \pm 0.55$ &1 \\
NGC 7679        &A &$42.74 \pm 0.12$   &$42.03 \pm 0.67$ &1 \\
\hline
UGC 2608        &N &43.63 ($<43.84$)$^{b}$&$43.56 \pm 0.22$ &4 \\
NGC 5135        &N &$43.24 \pm 0.08$   &$43.26 \pm 0.28$ &4
\enddata
\tablecomments{Columns: 
(1) Object name;
(2) merger stage (N = nonmerger, A = pre-merger, B = early-stage merger,
C = mid-stage merger, and D = late-stage merger; see \citealt{Yamada2019});
(3) nuclear 12~$\mu$m luminosity in erg s$^{-1}$;
(4) intrinsic 2--10 keV luminosity in erg s$^{-1}$;
(5) references for the intrinsic 2--10 keV luminosity.
List of references:
(1) \citet{Asmus2015};
(2) \citet{Oda2017};
(3) \citet{Romero2017};
(4) this work.}
\tablenotetext{a}{Typical error is adopted.}
\tablenotetext{b}{This value is converted from the total 12 $\mu$m luminosity by multiplying the mid-IR AGN fraction.}
\end{deluxetable}

These results ($\lambda_{\rm Edd}\sim0.1$)
are in contrast to the AGNs in late-stage merging
U/LIRGs, which often have large Eddington ratios of $\gtrsim$0.5,
such as Mrk~463E ($\lambda_{\rm Edd} \sim$ 0.4--0.8; \citealt{Yamada2018}), 
IRAS 05189--2524 ($\lambda_{\rm Edd} \sim 1.2$; \citealt{Teng2015}), 
and Mrk~231 ($\lambda_{\rm Edd} \sim 5.2$; \citealt{Teng2015}).
Thus, the SMBHs in these ``non-merging'' U/LIRGs are not 
so rapidly growing as those in typical ``merging'' U/LIRGs.

\begin{figure}
    \epsscale{1.11}
    \plotone{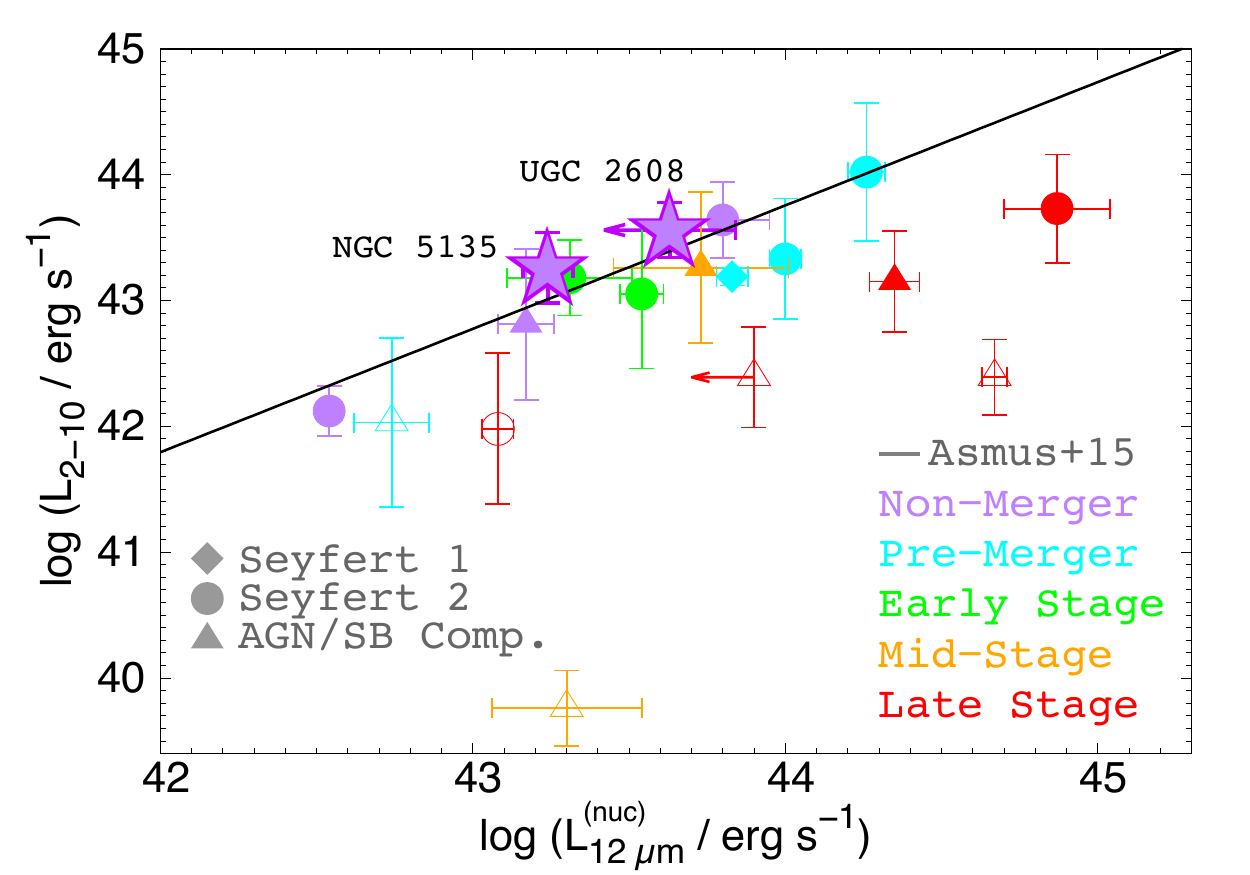}
    \caption{Nuclear 12 $\mu$m luminosity vs. intrinsic 2--10~keV luminosity for the local U/LIRGs in Table~\ref{T_12um}.
    These symbols are color coded by the merger stages determined by \citet{Stierwalt2013}.
    Diamonds, circles, and triangles mark the Seyfert 1/1.5, Seyfert 1.8/2, and AGN/starburst composites determined by the optical AGN classification, and empty symbols are the starburst-dominated objects whose mid-IR AGN fractions are $<$1/3 \citep[see][]{Yamada2019}. 
    Purple stars represent UGC 2608 and NGC 5135.
    Arrows mark upper limits, and black solid line shows the typical relation for local AGNs \citep{Asmus2015}.\\}
\label{F_12um_X}
\end{figure}

\subsection{Geometry of Narrow Line Region and Torus}
\label{Sub_geometry}

The comparison between intrinsic X-ray and nuclear 
(subarcsecond-scale) 12~$\mu$m luminosities is useful to
investigate the environment in the central regions. 
Nuclear 12~$\mu$m luminosity originates 
mainly from hot dust heated by an AGN \citep[e.g.,][]{Gandhi2009},
which are obtained from high spatial-resolution observations 
in order to minimize the contamination by the host galaxy
\citep{Asmus2014}. In fact, \citet{Asmus2015} found that
the nuclear 12~$\mu$m and intrinsic X-ray luminosities were well correlated 
for normal Seyferts,
whereas they were not for AGN/starburst composites, uncertain AGNs (e.g.,
low-luminosity AGNs), and buried AGNs with large torus covering
factors \citep{Yamada2019}. 

The nuclear 12~$\mu$m luminosities are obtained by \citet{Asmus2014} 
for 23 U/LIRGs in the GOALS sample.
In Table~\ref{T_12um}, we select AGNs in various merging
stages of these U/LIRGs \citep{Yamada2019} 
whose intrinsic 2--10 keV luminosities are estimated.
Figure~\ref{F_12um_X} shows the relation between X-ray and nuclear 12~$\mu$m
luminosities. As shown in the figure,
in the case of mid-/late-stage mergers, the X-ray to nuclear 12~$\mu$m
luminosity ratios are smaller than the typical relation found for normal Seyferts
\citep{Asmus2015}. There are two possibilities that (1) the
X-ray-to-bolometric luminosity ratio is small 
because of high Eddington ratios \citep{Vasudevan2007}
and that (2) starburst contamination to the 12 $\mu$m
luminosity is significant even in the nuclear region.  Whereas, UGC~2608\footnote{For
UGC~2608, the ``nuclear'' 12 $\mu$m luminosity is not available. Hence, we utilize
the total 12 $\mu$m luminosity [$(6.97 \pm 0.04)
\times 10^{43}$ erg s$^{-1}$ obtained from the 
\textit{Spitzer}/IRS spectra as a maximum value] by multiplying the
mid-IR AGN fraction ($\sim$62\%; \citealt{Diaz-Santos2017}).} and
NGC~5135 show the ratios consistent with the typical relation found for normal
Seyferts, indicating that they are neither AGNs with large Eddington
ratios nor starburst-dominant objects in agreement with their SFRs.

The properties of the tori in non-merging LIRGs are still
unclear. The IRAS 25--60~$\mu$m flux ratios ($f_{25}/f_{60}$) of 0.43 
for UGC~2608 and 0.34 for NGC~5135 \citep{Sanders2003} show warm far-IR 
colors ($>$0.2), supporting that they are not buried AGNs 
but normal edge-on CT AGNs with co-existing nuclear starbursts
\citep{Sanders1988,Imanishi2006,Imanishi2007}.
To confirm the torus geometry, the [\ion{O}{4}] 
25.89~$\mu$m and X-ray luminosity ratio provides the information of 
the spatial extent
of the narrow line region (NLR).
The [\ion{O}{4}] line is emitted from the NLR
irradiated by UV light from an AGN. Hence, its luminosity should be
roughly proportional to the AGN bolometric luminosity times the solid 
angle of the NLR \citep[e.g.,][]{Kawamuro2016}. 
However, this relation is also affected by the 
ratio of the bolometric luminosity to the X-ray one;
for instance, AGNs with high Eddington ratio such as those
in the late-stage merging U/LIRGs may 
have large ratios (i.e., X-ray weak; e.g., 
\citealt{Teng2015,Yamada2018,Toba2019}).

\begin{figure}
    \epsscale{1.11}
    \plotone{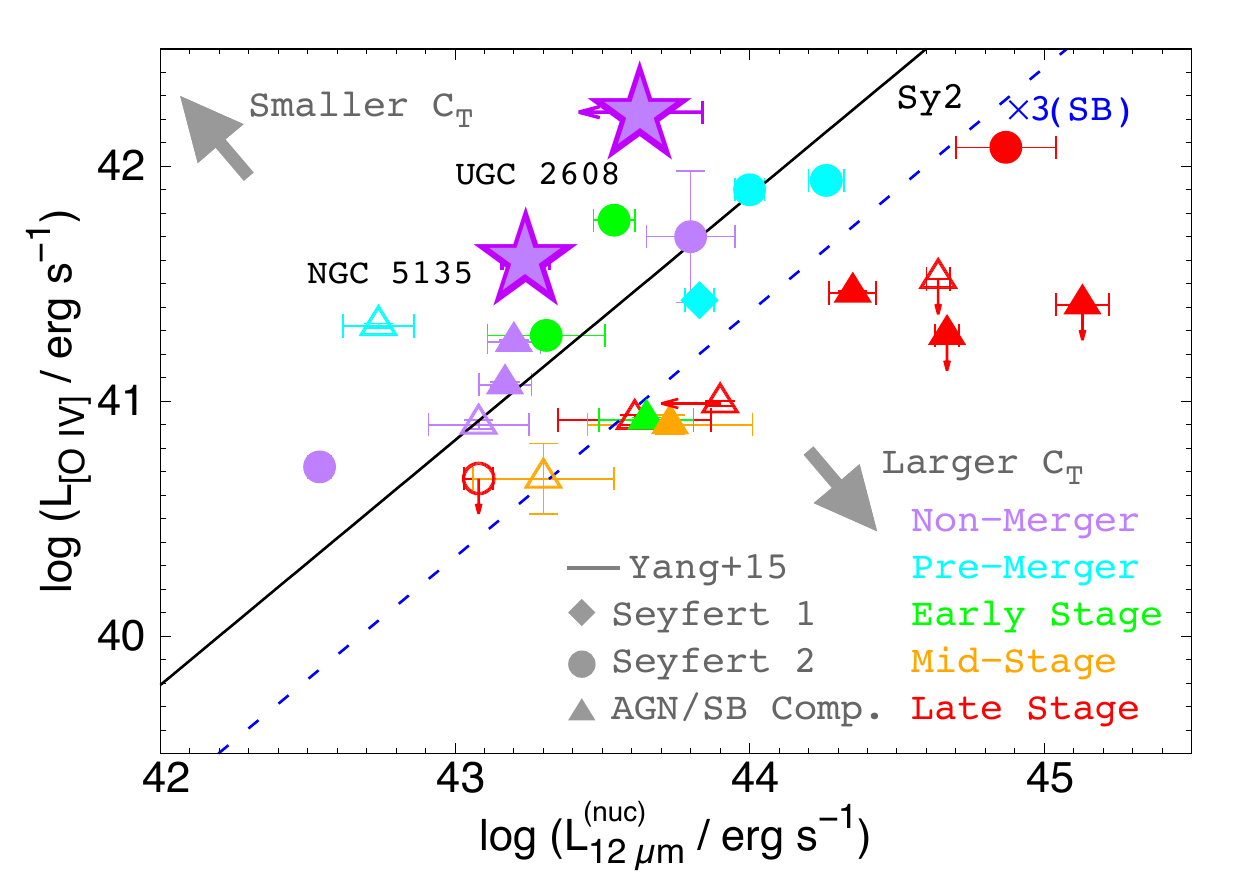}
    \caption{Relation of the [\ion{O}{4}] 25.89 $\mu$m to nuclear
 12~$\mu$m luminosity for our targets and the AGNs in the local U/LIRGs
 \citep{Yamada2019}. Symbols are 
the same as in Figure~\ref{F_12um_X}.
    The black solid and blue dashed lines are the averaged relation for
 Seyfert~2s obtained by \citet{Yang2015} and that corrected for
 contribution of starburst in the mid-IR luminosity by a factor of 3,
which is valid 
for objects with mid-IR AGN fractions of 1/3 
(a typical value in starburst-dominant objects).\\}
\label{F_12um_oiv}
\end{figure}

To avoid the uncertainty by this effect, we investigate
the correlation between the 
[\ion{O}{4}]~25.89~$\mu$m line and nuclear 12~$\mu$m continuum luminosities.
This would be a better tracer of covering factors 
because both are determined mainly by the UV luminosity dominating
the total power of the intrinsic AGN radiation
\citep{Yamada2019}.
The sample is selected from our targets and the 
local U/LIRGs listed in Table~2 of \citet{Yamada2019}, 
whose [\ion{O}{4}] and nuclear 12~$\mu$m luminosities 
are both available.
The result is shown in Figure~\ref{F_12um_oiv}. 
They reported that the AGNs in the mid- to late-stage mergers
of U/LIRGs have small [\ion{O}{4}] luminosities relative to the 12~$\mu$m
ones, suggesting that their nuclei are deeply buried by tori with large
covering factors. By contrast, the AGNs in UGC~2608 and NGC~5135 have
larger [\ion{O}{4}] luminosities than those of typical Seyfert~2s
\citep{Yang2015} and the other U/LIRGs, indicating that 
the solid angles of their NLRs are large 
(i.e., the torus covering factors are small).

\begin{figure*}
    \epsscale{1.11}
    \plottwo{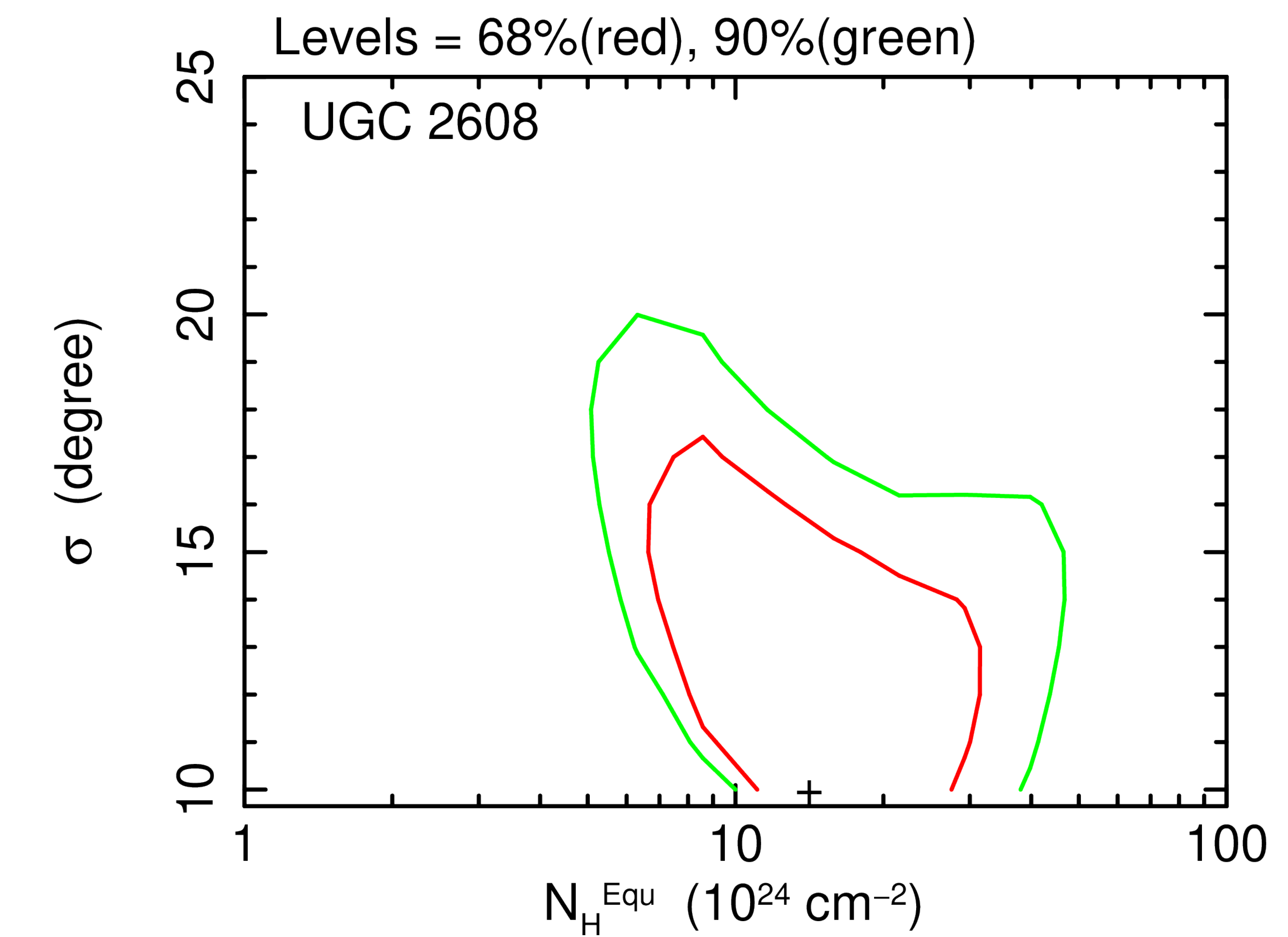}{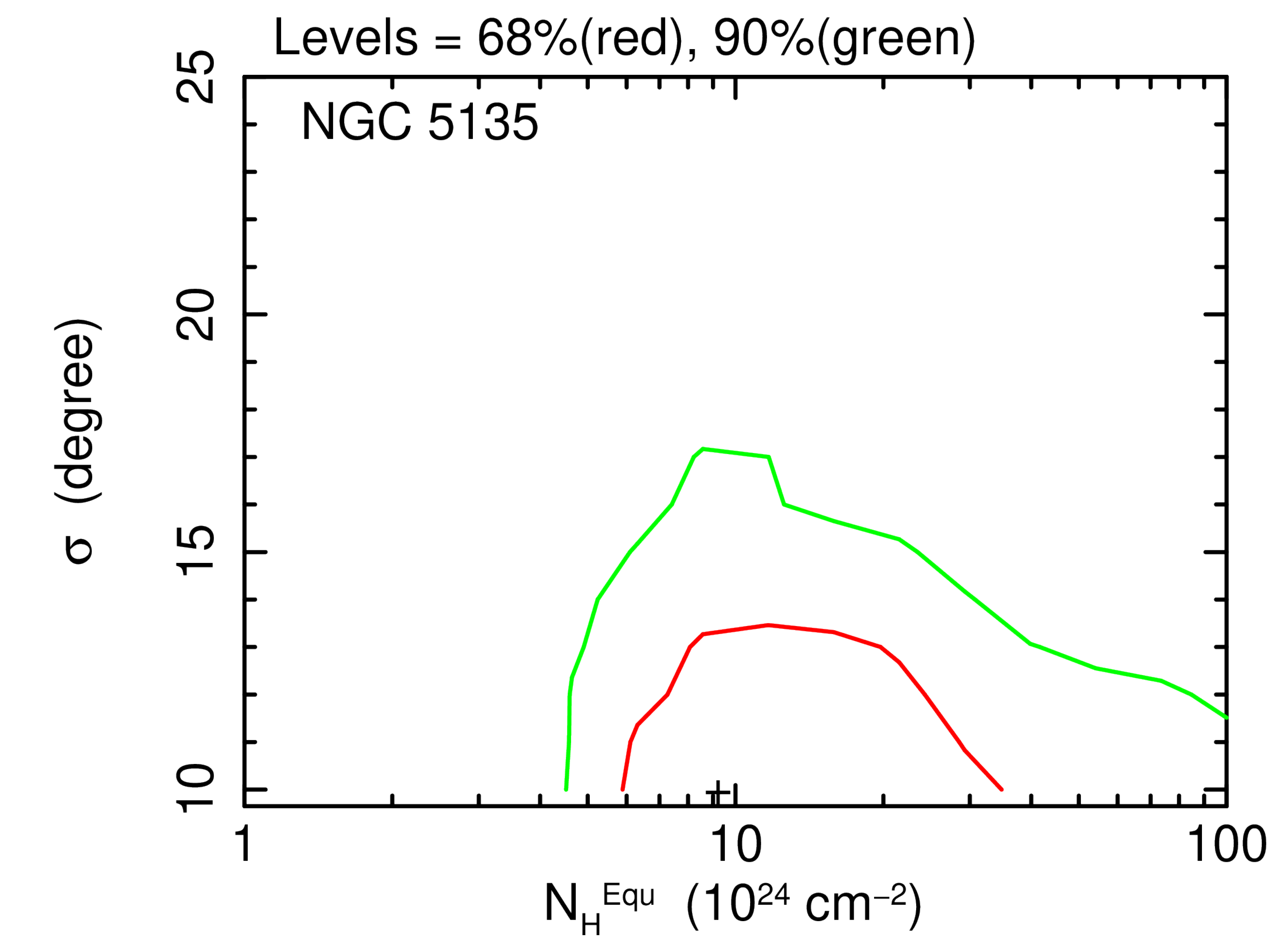}
    \caption{Confidence contours between $N_{\rm H}^{\rm Equ}$ and $\sigma$ of UGC 2608 (left panel) and NGC 5135 (right panel).
    The red and green lines show the confidence levels at 1$\sigma$ (red) and 90\% (green), respectively. \\}
\label{F_contour}
\end{figure*}

\subsection{Covering factors in the X-ray and IR band}
\label{Sub_irvsX-ray}
To reinforce our interpretation on the torus geometry of UGC~2608 and NGC~5135,
we estimate the covering factor ($C_{\rm T}$) from our X-ray spectral
analysis results. With the XCLUMPY
model, we have determined the torus angular widths to be 
$\sigma <$ 18.2\degr~degrees and $<$14.5\degr,
and the equatorial hydrogen column densities to be 
$N_{\rm H}^{\rm Equ} = 15^{+19}_{-9}$ 
$\times 10^{24}$~cm$^{-2}$ and
$9.5^{+32.0}_{-3.9}$ $\times 10^{24}$~cm$^{-2}$ 
for UGC~2608 and NGC~5135, respectively.
These torus angular widths are smaller than the averaged
value of the 10 local Seyfert~2s (\textless$\sigma$\textgreater\ $\sim 30\degr$) 
analyzed by \citet{Tanimoto2020} with XCLUMPY.
Figure~\ref{F_contour} shows the confidence contours between 
$N_{\rm H}^{\rm Equ}$ and $\sigma$.
In the geometry of XCLUMPY, 
the mean column density at a given elevation angle ($\theta \equiv
90\degr - i$) can be calculated as
\begin{align}
&N_{\rm H}(\theta) = N_{\rm H}^{\rm Equ}{\rm exp}\left[-\left(\frac{\theta}{\sigma}\right)^2\right].
\end{align}
Defining $\theta_{\rm c}$ as the angle for which
$N_{\rm H} = 10^{22}$ cm$^{-2}$, we find
$\theta_{\rm c} < 49.4\degr$ and
$<$39.6\degr, which correspond to $C_{\rm T} < 0.76$ and 
$<$0.64 for UGC~2608 and
NGC~5135, respectively. These results imply that 
these AGNs are not deeply buried. 

Considering the mid-IR results in Section~\ref{Sub_geometry},
the AGNs in UGC~2608 and NGC~5135 have small covering factors
for both gas and dust. 
\citet{Ichikawa2015} analyzed
the IR SED of NGC 5135
with the CLUMPY model and obtained
a torus angular width of
$\sigma_{\rm IR} = 63^{+3}_{-5}$ degrees 
and a covering factor of 
$C_{\rm T,IR} = 0.97^{+0.01}_{-0.04}$ (the errors are 1$\sigma$).
These estimates are larger than those of X-ray results 
($\sigma < 14.5\degr$ and $C_{\rm T} < 0.76$).
Similar trends are reported in case of local Seyfert 1s
\citep{Ogawa2019} and Seyfert 2s \citep{Tanimoto2019,Tanimoto2020}
by comparing the results with the XCLUMPY and CLUMPY models.
\citet{Tanimoto2020} and \citet{Ogawa2020} find a 
significant tendency that $\sigma_{\rm IR}$ estimated by 
\citet{Ichikawa2015} and \citet{Garcia-Bernete2019} is
larger than $\sigma_{\rm X}$, particularly in a high
inclination system. They propose that the discrepancy
can be explained by a contribution from
``dusty'' polar outflows in the IR flux
\citep{Tristram2014,Asmus2016,Asmus2019}. These effects
are not considered in the CLUMPY model,
whereas the effects to the X-ray spectra are expected to be 
small \citep{Tanimoto2020}.
Thus, the results from the X-ray spectra with XCLUMPY
give more realistic estimates of the torus covering factors
than those from the IR SED with CLUMPY.

Even if the nuclear 12~$\mu$m luminosity ratio 
may contain the emission from the polar outflows, 
the [\ion{O}{4}] to nuclear 12~$\mu$m luminosity ratio
still represents the spatial extent of the NLR. 
Hence, we conclude that AGN tori in UGC 2608 and NGC 5135 are 
geometrically thin, on the basis of the X-ray spectra and 
the [\ion{O}{4}] to nuclear 12~$\mu$m luminosity ratios.

\subsection{Covering Factors vs. Eddington ratios}
\label{Sub_ctvsEddington}
We investigate the relation between the covering factor
and Eddington ratio. Utilizing local AGNs in the \textit{Swift}/BAT
70-month catalog, \citet{Ricci2017cNature} report that AGNs with
$10^{-4} \leq \lambda_{\rm Edd} \leq 10^{-1.5}$ have
obscurers of $N_{\rm H} \geq 10^{22}$ cm$^{-2}$
with large covering
factors ($C_{\rm T} \sim 0.85$), whereas AGNs with $\lambda_{\rm Edd}
\geq 10^{-1.5}$ have those with 
smaller covering factors ($C_{\rm T} \sim 0.40$).
They interpret that radiation pressure from the AGN 
expells dusty material when $\lambda_{\rm Edd} \geq 10^{-1.5}$. 

By adopting the estimated Eddington ratios (Section~\ref{Sub_bh}),
the relation of \citet{Ricci2017cNature} predicts  
the covering factors with $N_{\rm H} \geq 10^{22}$~cm$^{-2}$
to be $C_{\rm T} =$ 0.32--0.76 
(UGC 2608) and 0.31--0.72 (NGC 5135), and that
with $N_{\rm H} \geq 10^{24}$~cm$^{-2}$ to be $C_{\rm T}^{(24)} \sim 0.22$
(both objects).\footnote{The 1$\sigma$ uncertainties 
are $\pm$0.08 (UGC~2608) and $\pm$0.09 (NGC~5135) for $C_{\rm T}$, 
and $\pm$0.07 (both) for $C_{\rm T}^{\rm (24)}$.}
These values match well with the obtained covering factors 
in the XCLUMPY model ($C_{\rm T} \sim 0.45$ and $\sim$0.44, and 
$C_{\rm T}^{(24)} \sim 0.28$ and $\sim$0.26 
for UGC~2608 and NGC~5135, respectively). 
The results suggest that the XCLUMPY model 
reproduces a realistic description of their compact AGN tori 
with $N_{\rm H} \sim 10^{24}$~cm$^{-2}$ obscuration; 
foreground absorption by the host galaxy
(typically $N_{\rm H} \sim 10^{22}$~cm$^{-2}$), if any, 
is negligible. We infer that 
their tori become geometrically thin due to 
significant radiation pressure of the AGN as is the case of 
local AGNs with moderately high Eddington ratios.

Thus, our results indicate that the CT AGNs in 
these ``non-merging'' LIRGs
are just a normal AGN population with moderate Eddington ratios
seen close to edge-on through a large
line-of-sight column density.
We note that these two objects may have smaller covering factors than
typical values in non-merging LIRGs because they have largest
[\ion{O}{4}]-to-12~$\mu$m luminosity ratios among the 24 local U/LIRGs in
Figure~\ref{F_12um_oiv}. This may be due to a selection bias 
by the detections with \textit{NuSTAR} for 
AGNs with high luminosities and hence with large Eddington ratios. 
It is thus important to confirm if
all non-merging LIRGs follow the relation by 
\citep{Ricci2017cNature}, using a larger sample
by more sensitive broadband X-ray spectroscopy.

By contrast, many studies suggest that AGNs in U/LIRGs of late-stage mergers
are deeply buried (i.e., 
having tori with large covering factors reaching almost unity; 
\citealt{Imanishi2006,Imanishi2007,Ricci2017bMNRAS,Yamada2019})
even if they shine at large Eddington ratios \citep{Teng2015,Oda2017,Yamada2018}. 
Then, the driver of these phenomena 
would be not the Eddington ratio but the merger environment.
In such a ``merging'' system, 
the column density in almost all directions becomes too large to be 
blowed out by radiation pressure of the AGN.
Thus, we suggest that AGNs in ``non-merging'' LIRGs and those in ``merging'' 
ones are distinct populations. 
The former have small to moderate Eddington ratios and are not buried, 
while the latter have moderate to high Eddington ratios 
even though they are deeply ``buried''.
This is probably because the AGN trigger mechanism in ``non-merging'' 
LIRGs is less violent than major mergers, 
such as minor mergers, fly-by companions, and/or secular evolution
\citep[e.g.,][]{Alonso-Herrero2013}.
\\

\section{Summary and Conclusions}
\label{S_conclusion}

In this paper, we study the best-quality broadband X-ray spectra
of the two ``non-merging'' LIRGs UGC~2608 and NGC~5135, 
by combining the
data of \textit{NuSTAR}, \textit{Suzaku}, \textit{XMM-Newton}, and
\textit{Chandra}. Applying the XCLUMPY model \citep{Tanimoto2019}, we
estimate the line-of-sight hydrogen column densities of $N_{\rm H}^{\rm
LOS}$ = $5.4^{+7.0}_{-3.1}$ $\times 10^{24}$~cm$^{-2}$ and
$6.6^{+22.5}_{-2.7}$ $\times 10^{24}$~cm$^{-2}$, and the intrinsic
2--10~keV luminosities to be $L_{\rm 2-10} =$ $3.9^{+2.2}_{-1.7}$ 
$\times 10^{43}$~erg~s$^{-1}$ and 
$2.0^{+3.3}_{-1.0}$ $\times 10^{43}$~erg~s$^{-1}$ for
UGC~2608 and NGC~5135, respectively. Thus, both objects are 
firmly classified as CT AGNs.

The Eddington ratios of both targets are estimated to be $\sim$0.1, by
using the black hole masses \citep{Dasyra2011,Marinucci2012} and
the intrinsic 2--10 keV luminosities obtained from our careful analysis. 
From the X-ray spectra, we determine the torus angular widths of 
UGC~2608 and NGC~5135 to be
$\sigma < 20\degr$, and the covering factors of material with
$N_{\rm H}>10^{22}$~cm$^{-2}$ to be $C_{\rm T} \lesssim$ 0.7.
The covering factors and Eddington ratios 
are consistent with the
relation found for local \textit{Swift}/BAT AGNs by \citet{Ricci2017cNature},
implying that their tori are geometrically thin because of radiation
pressure of the AGN to dusty material.

Our results indicate that the CT AGNs in these
``non-merging'' LIRGs are just a normal AGN population with moderate
Eddington ratios seen close to edge-on through a large line-of-sight
column density. 
By contrast, AGNs in U/LIRGs of late-stage mergers 
are deeply buried \citep{Imanishi2006,Ishisaki2007,Ricci2017bMNRAS,Yamada2019} 
despite of their large Eddington ratios 
\citep{Teng2015,Oda2017,Yamada2018}. 
This suggests that AGNs in ``non-merging'' LIRGs and those
in ``merging'' ones are distinct populations.

\acknowledgments
We thank the anonymous referee for helpful suggestions, which helped us improve the original manuscript.
This research has made use of the \textit{NuSTAR} Data Analysis Software
(NuSTARDAS) jointly developed by the ASI Science Data Center (ASDC,
Italy) and the California Institute of Technology (Caltech, USA).  This
work makes use of data obtained from the \textit{Suzaku} satellite, a
collaborative mission between the space agencies of Japan (JAXA) and the
USA (NASA).  This publication makes use of data obtained with
\textit{XMM-Newton}, an ESA science mission with instruments and
contributions directly funded by ESA Member States and NASA, and with
\textit{Chandra}, supported by the \textit{Chandra} X-ray Observatory
Center, which is operated by the Smithsonian Astrophysical Observatory
for and on behalf of NASA.  This publication also makes use of the data
from the NASA/IPAC Infrared Science Archive and NASA/IPAC Extragalactic
Database (NED), which are operated by the Jet Propulsion Laboratory,
California Institute of Technology, under contract with the National
Aeronautics and Space Administration.

This work is financially supported by the Grant-in-Aid for Scientific
Research 19J22216 (S.Y.), 17K05384 and 20H01946 (Y.U.), 17J06407
(A.T.), 15K05030 (M.I.), 18J01050 and 19K14759 (Y.T.).
C.R. acknowledges the support from the CONICYT+PAI Convocatoria Nacional subvencion a
instalacion en la academia convocatoria a\~{n}o 2017 PAI77170080.

\vspace{5mm}
\facilities{\textit{NuSTAR}, \textit{Suzaku}, \textit{XMM-Newton}, \textit{Chandra}}


\software{XCLUMPY \citep{Tanimoto2019}, HEAsoft 6.25, XSPEC \citep{Arnaud1996}, NuSTARDAS, SAS 17.00 \citep{Gabriel2004}}

\bibliographystyle{aasjournal.bst}
\bibliography{reference}

\end{document}